\documentclass[pra,aps,twocolumn,amsmath,amssymb,superscriptaddress]{revtex4-1}
\usepackage{epsfig,amsmath}
\usepackage{subfigure}
\usepackage{graphicx}
\usepackage{dcolumn}
\usepackage{stmaryrd}
\usepackage{mathrsfs}
\usepackage{pifont}
\usepackage{amsthm}
\usepackage{amssymb}
\usepackage{bm}
\usepackage{latexsym}
\usepackage[colorlinks=true,linkcolor=blue,citecolor=blue]{hyperref}
\usepackage{color}
\usepackage{epstopdf}
\usepackage[ruled]{algorithm2e}
\usepackage{lipsum}

\begin{document}

\title{Generating Fock states over 10000 excitations with near-unit fidelity via adaptive generalized-parity measurement}

\author{Chen-yi Zhang}
\affiliation{School of Physics, Zhejiang University, Hangzhou 310027, Zhejiang, China}

\author{Jun Jing}
\email{Contact author: jingjun@zju.edu.cn}
\affiliation{School of Physics, Zhejiang University, Hangzhou 310027, Zhejiang, China}

\date{\today}

\begin{abstract}
Fock states are fundamental quantum states with a precisely defined integer number of excitations, serving as the core basis for describing bosonic modes. Large Fock states provide irreplaceable non-classical resources for quantum information processing and quantum metrology. The deterministic generation of macroscopic photon-number Fock states has long been a difficult problem in the field of quantum optics. We propose an adaptive generalized parity measurement (GPM) protocol for generating Fock states with more than $10000$ excitations, avoiding low success probability subject to postselection and high cost under complex coherent controls. For general discrete-spectrum systems, e.g., a bosonic mode coupled to an ancillary qubit, we derive a construction rule in which the intervals between repeated measurements on qubit are updated adaptively based on the last outcome. It means that our protocol does not discard any measurement trajectory, dramatically different from the probabilistic protocols that retain only one prescribed trajectory of postselection. In the resonant Jaynes-Cummings model, a large coherent state can be almost deterministically transformed to a large Fock state of $n_t=\mathcal{O}(10^4)$ excitations by $10$ rounds of measurements, the average fidelity of which is about $87\%$. The success probability for obtaining $|20000-\sqrt{20000}\leq n_t\leq20000+\sqrt{20000}\rangle$ with a fidelity above $99\%$ is about $35\%$ with respect to the ensemble sampling. Our protocol is fault-tolerant in the presence of moderate measurement error and parametric imperfection. Also it remains effective when the system is prepared as displaced thermal states, showing reliable performance regardless of initial state.
\end{abstract}

\maketitle

\section{Introduction}

Bosonic modes are fundamental elements of quantum information processing~\cite{RMPBraunstein2005,RMPWeedbrook2012} over a broad range of platforms including cavity and circuit quantum electrodynamics (QED)~\cite{PRLPeaudecerf2014,NatrueGuerlin2007,NatrueDeleglise2008,RMPBlais2021}, trapped ions~\cite{PRLMeekhof1996,PRLBenKish2003,PRLMatsos2024,PRXQuantumValahu2024}, optomechanical systems~\cite{RMPAspelmeyer2014}, and hybrid magnonic systems~\cite{PhysRepZare2022}. Engineering quantum states of bosonic modes is therefore a central task in quantum science and technology. In the nonclassical states, Fock states provide a rigorous framework for the second quantization formulation of quantum mechanics and a foundational significance for quantum metrology~\cite{NatPhysDeng2024}, quantum computation~\cite{PRLLloyd1999,PRXMichael2016,PRLHeeres2015}, and quantum communication~\cite{RMPCaves1994,NatureBouwmeester1997}. Large Fock states hold irreplaceable value in advancing quantum simulation~\cite{npjQISturges2021}, exploring macroscopic quantum features~\cite{RMPFrowis2018}, and enhancing quantum sensing~\cite{PRLHolland1993,PRAUys2007}. In a recent demonstration, microwave Fock states approaching $100$ photons obtained a displacement-sensing gain of $14.8\pm0.2$ dB and a phase-sensing gain of $12.3\pm0.5$ dB~\cite{NatPhysDeng2024} over the standard quantum limit.

Protocols for generating Fock states have been presented in cavity QED~\cite{PRLBrune1990,PRLVogel1993,PRLLaw1996,PRLBrattke2001,PRLFranifmmode2001,
NatureSayrin2011,PRLUria2020,PRLZhou2012}, circuit QED~\cite{NatureHofheinz2008,NatCommunPremaratne2017,PRLWang2017}, and mechanical oscillator~\cite{PRATan2014,NatureChu2018} systems, but most of them are restricted to a small number of excitations. Scaling up the excitation number to the macroscopic regime remains challenging. Resonant subspace engineering was employed to confine the bosonic dynamics to an invariant subspace spanned by the initial coherent state and the target Fock state, enabling a large-Fock-state generation up to $70$-$100$ photon numbers in $5$ iterations of operations~\cite{arxivJin2026}. To overcome the scalability bottleneck, Kerr-engineered Fock-space lenses have been used to focus an initial coherent-state distribution into Fock states of $10000$ photons with fidelity up to $73\%$ in a circuit QED platform~\cite{arxivLi2026}. A similar technique that combines optimized evolution and displacement is proposed to generate Fock states with photon number up to $N\approx100$ with fidelities about $90\%$~\cite{Xiong2026LargeFock}. However, these protocols are generally subject to the pure-state initialization and demanding multiple coherent controls, including precisely tailored external driving, gate operations, and state manipulation.

Quantum state engineering by virtue of measurements on a coupled ancilla offers an alternative strategy with little demand on precise state initialization~\cite{JMP1977Misra,PRL2003Nakazato,PRANakazato2004,PRAYan2023}. Instead of coherently transferring population directly to the target state, repeated projective measurements on the ancilla progressively filter out unwanted components and drive the main system toward the desired state. This mechanism has been applied to ground-state preparation~\cite{PRBLi2011,PRAYan2022,PRAKonar2022,PRAJin2022}, quantum-battery charging~\cite{PRAppliedYan2023,PRATinggui2024}, and entanglement generation~\cite{PRAWu2004}. Measurement-based generation of large Fock states has been investigated with both dispersive~\cite{NatPhysDeng2024,NatureSayrin2011,PRLZhou2012,PRLWang2017} and resonant~\cite{PRAZhang2024,PRAZhang2026} qubit-resonator couplings. Conventional protocols, however, have to select a single prescribed measurement trajectory and discard all the others. The relevant experimental cost grows rapidly with the measurement number, while the ultimate success probability is limited by the initial population of the target Fock component. From a coherent state $|\alpha\rangle$ centered around the target Fock state $|n_t\approx|\alpha|^2\rangle$, the success probability $P=e^{-|\alpha|^2}|\alpha|^{2n_t}/n_t!$ becomes lower than $1\%$ when $n_t>1500$. It severely limit the scalability of postselection-based protocols.

Adaptive measurement provides a natural protocol to overcome the bottleneck in success probability~\cite{PRLWiseman1995,PRLBerryWiseman2000}, by which the parametric setting at each step is updated according to the last measurement record and all possible records are incorporated. It has been studied in quantum metrology~\cite{PRLArmen2002,PRATsang2008,PRLWheatley2010,ScienceYonezawa2012,PRLSalvia2023} and quantum state preparation~\cite{NJPYuasa2009,PRATanaka2012,NatureRiste2013,PRAWang2021,PRXQuantumSmith2024,PRLYu2026}. We develop a general adaptive rule for deterministically constructing generalized parity measurements (GPM) in systems of discrete and nondegenerate spectra. The target system is coupled to an ancillary qubit through an exchange-type interaction; and the free-evolution duration before each qubit-measurement is updated due to the last measurement outcome. Besides success probability, it takes advantage of the efficiency of GPM in the number of measurements, showing a logarithmic scaling similar to the number of ancillary qubits in quantum phase estimation~\cite{NielsenChuang2010} or the iteration rounds of a Ramsey-type circuit~\cite{PRL2014Asadian,PRX2018Fluhmann,NatPhysDeng2024,PRAZhang2026}. No external driving~\cite{PRLYu2026}, gate operations~\cite{PRAWang2021}, or changes of measurement basis~\cite{PRATanaka2012} are required. In the Jaynes-Cummings (JC) model, our protocol can generate a Fock state with $n_t=20000$ within only $10$ measurement rounds. Its feasibility in realistic implementations is supported by the robustness against moderate readout-error and detuning deviation. Our protocol also works for mixed states such as displaced thermal states, superior to the unitary-operation methods that generally assume pure state initialization. Significant success probability~\cite{NatPhysDeng2024,PRAZhang2026}, low-cost in initialization and external operations~\cite{arxivJin2026,arxivLi2026}, and high efficiency~\cite{PRAZhang2024} therefore establish the importance of our protocol for quantum state engineering by measurement.

The rest of this paper is organized as follows. We introduce a general theory in Sec.~\ref{gene_model} for quantum state engineering by measurements on an ancillary qubit coupled to the main system by an exchange-like interaction. In Sec.~\ref{adap_rule}, we develop a deterministic protocol about the preceding engineering for constructing adaptive generalized parity measurements on the main system with respect to any basis states determined by the system coupling operator. The induction proof about the relevant adaptive rule for the measurement intervals can be found in Appendix~\ref{Proof}. In Sec.~\ref{ideal}, we apply the protocol to the JC model to generate macroscopic Fock states. Then we discuss its fault-tolerance with readout errors and thermal environment in Secs.~\ref{error} and \ref{thermal}, respectively. We summarize our results in Sec.~\ref{conclusion}.

\section{Theoretical framework}

\subsection{General model for a composite system}\label{gene_model}

For realizing the generalized parity measurements in a high-dimensional and nondegenerate system through repeated projective measurements on a coupled ancillary qubit, the Hamiltonian of the composite system reads
\begin{equation}\label{H_total}
H=\frac{\Delta}{2}\sigma_z+g\left(Q\sigma_++Q^\dagger\sigma_-\right)
=\begin{bmatrix}
\frac{\Delta}{2} I~&gQ\\
gQ^\dagger~& -\frac{\Delta}{2}I
\end{bmatrix},
\end{equation}
where $\Delta$ denotes the detuning between the ancillary qubit and the main system and $g$ is their coupling strength. $Q$ and $I$ represent the coupling operator and the identity operator living in the space of the main system, respectively. The raising and lowering operators of the qubit are defined as $\sigma_+=|e\rangle\langle g|$ and $\sigma_-=|g\rangle\langle e|$, respectively.

After a duration $\tau$, the joint time-evolution operator associated with Eq.~(\ref{H_total}) can be written as
\begin{equation}\label{U_total}
U(\tau)=e^{-iH\tau}=\sum_{n=0}^\infty\frac{(-i\tau)^nH^n}{n!}=\begin{bmatrix}
    V_{ee}(\tau)~&V_{eg}(\tau)\\
    V_{ge}(\tau)~&V_{gg}(\tau)
\end{bmatrix},
\end{equation}
where the Kraus operators read
\begin{subequations}\label{Kraus}
\begin{align}
V_{ee}(\tau)&\equiv \cos(\Omega\tau)-i\frac{\Delta}{2}\Omega^{-1}\sin(\Omega\tau),\\
V_{gg}(\tau)&\equiv \cos(\Lambda\tau)+i\frac{\Delta}{2}\Lambda^{-1}\sin(\Lambda\tau),\\
V_{ge}(\tau)&\equiv -igQ^\dagger\Omega^{-1}\sin(\Omega\tau),\\
V_{eg}(\tau)&\equiv -igQ\Lambda^{-1}\sin(\Lambda\tau),
\end{align}
\end{subequations}
with $\Omega\equiv\sqrt{g^2QQ^\dagger+I\Delta^2/4}$ and $\Lambda\equiv\sqrt{g^2Q^\dagger Q+I\Delta^2/4}$. According to Naimark's dilation theorem~\cite{JPAPellonpaa2023}, a projection on a subsystem of a composite system generally yields a positive-operator-valued measure (POVM) on the rest part. When the ancillary qubit is initially in the state $|j\rangle$ and the main system is prepared as $\rho_c$, a projective measurement $M_i=|i\rangle\langle i|$ on the qubit after the joint evolution in Eq.~(\ref{U_total}) yields a POVM on the main system with the Kraus operator $V_{ij}(\tau)=\langle i|U(\tau)|j\rangle$. Here $|i\rangle$ and $|j\rangle$ are arbitrary normalized states of the qubit. After a round of free-evolution and measurement, the density operator of the main system becomes $\rho_c'=V_{ij}(\tau)\rho_cV_{ij}^\dagger(\tau)/P_{ij}$ with a success probability $P_{ij}={\rm Tr}[V_{ij}(\tau)\rho_cV_{ij}^\dagger(\tau)]$.

\subsection{Adaptive rule for deterministically realizing generalized parity measurements}\label{adap_rule}

We assume that $Q^\dagger$ possesses a ``ladder'' structure
\begin{equation}\label{Qdagger}
Q^\dagger=\sum_m \mathcal{Q}_m|\lambda_{m+1}\rangle\langle \lambda_m|,
\end{equation}
where $\mathcal{Q}_m$ is an arbitrary complex number and $\{|\lambda_m\rangle\}$'s constitute a completed and orthogonal set of bases with $\langle\lambda_m|\lambda_k\rangle=\delta_{mk}$. $\Omega$ and $\Lambda$ in Eq.~(\ref{Kraus}) are therefore diagonal in this basis and can be written as
\begin{equation}\label{OmegaLambda}
\Omega=\sum_m\Omega_m|\lambda_m\rangle\langle\lambda_m|, \quad
\Lambda=\sum_m\Omega_{m-1}|\lambda_m\rangle\langle\lambda_m|
\end{equation}
with $\Omega_m=\sqrt{g^2|\mathcal{Q}_m|^2+\Delta^2/4}$, respectively. Consequently, the Kraus operators in Eqs.~(\ref{Kraus}) can take the form
\begin{subequations}
\begin{align}\label{Vee}
V_{ee}(\tau)&=\sum_m\alpha_m(\tau)|\lambda_m\rangle\langle \lambda_m|,\\ \label{Vgg}
V_{gg}(\tau)&=\sum_{m}\alpha^*_{m-1}(\tau)|\lambda_m\rangle\langle \lambda_m|,\\ \label{Vge}
V_{ge}(\tau)&=-i\sum_m\beta_m(\tau)|\lambda_{m+1}\rangle\langle\lambda_m|,\\ \label{Veg}
V_{eg}(\tau)&=-i\sum_m\beta^*_{m-1}(\tau)|\lambda_{m-1}\rangle\langle\lambda_{m}|
\end{align}
\end{subequations}
with coefficients
\begin{subequations}\label{alphabeta}
\begin{align}
\alpha_m(\tau)&=\cos(\Omega_m\tau)-i\frac{\Delta}{2\Omega_m}\sin(\Omega_m\tau),\label{alpham}\\
\beta_m(\tau)&=\frac{g\mathcal{Q}_m}{\Omega_m}\sin(\Omega_m\tau).\label{betam}
\end{align}
\end{subequations}
We further assume that the coefficient $\mathcal{Q}_m$ can be expressed by
\begin{equation}\label{Qk}
|\mathcal{Q}_m|^2=\mu n_m+\nu,
\end{equation}
where $n_m\in\mathbb{Z}$ depends on $m$, and $\mu$ and $\nu$ are constants irrespective of $m$. Thus, the Rabi frequency associated with the basis $|\lambda_m\rangle$ reads
\begin{equation}\label{Rabi_fre}
\Omega_m=\sqrt{g^2|\mathcal{Q}_m|^2+\Delta^2/4}=\sqrt{g^2\left(\mu n_m+\nu\right)+\Delta^2/4}.
\end{equation}

\begin{figure}[htbp]
\centering
\includegraphics[width=0.95\linewidth]{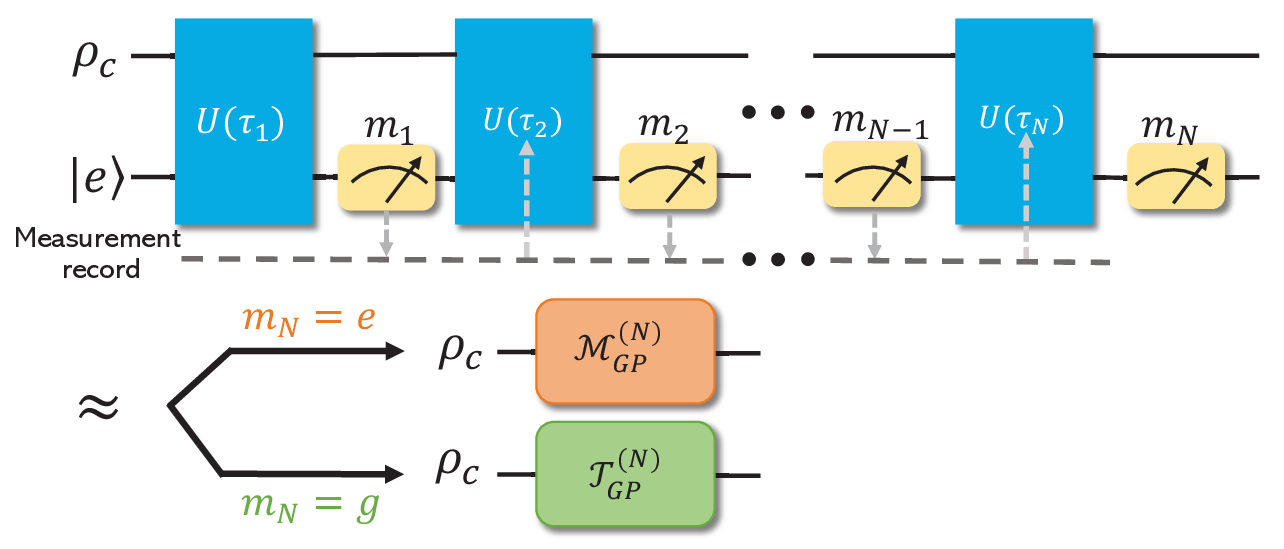}
\caption{Circuit diagram about constructing the adaptive generalized parity measurement by performing rounds of projective measurements on the ancillary qubit with outcomes $m_k=e$ or $g$. The intervals $\tau_k$ between neighboring measurements depend on the last measurement record. $N$ rounds of measurements yield a GPM $\mathcal{M}_{\rm GP}^{(N)}$ [a displaced GPM $\mathcal{T}_{\rm GP}^{(N)}$] on the main system $\rho_c$ if the final outcome is $m_N=e$ ($m_N=g$).}\label{Circuit_Diagram}
\end{figure}

Figure~\ref{Circuit_Diagram} is the diagram of the our protocol, in which the duration of free evolution is adaptively tuned by the last measurement outcome. We use a set of integers $n_t(k)$'s as GPM labels to denote the central component in the generalized-parity subspace after $k$ rounds of measurement. The initial GPM label $n_t(0)$ is typically set as the integer mostly close to the average population of the main system, which should be sufficiently large if the target is a basis state $|\lambda_m\rangle$ with a large index $m$ or a mixture of them. We also identify the generalized-parity (modulo $2^k$) subspace spanned by the basis states $\{|\lambda_m\rangle\}$ satisfying $n_m=n_t(k)\ {\rm mod}\ 2^k$ after $k$ rounds. $k=1$ describes the parity in the normal sense. The evolution duration of the $k$th round is set by
\begin{equation}\label{tauk}
\tau_k=l_kt_{n_t(k-1)},\quad t_{n_t(k-1)}=\frac{\pi}{\Omega_{n_t(k-1)}}
\end{equation}
with $l_k=\lceil [n_t(k-1)+\nu/\mu]/2^{k-1}\rceil $ or $\lfloor [n_t(k-1)+\nu/\mu]/2^{k-1}\rfloor$. And $\Omega_{n_t(k)}$ is the Rabi frequency in Eq.~(\ref{Rabi_fre}) with $n_m=n_t(k)$. In the nearby regimes for each $|n_t(k)\rangle$, we consider the basis states $|\lambda_m\rangle$'s that satisfy
\begin{equation}\label{condition}
|n_m-n_t(k)|\ll n_t(k)+\nu/\mu.
\end{equation}
Under resonant condition $\Delta=0$ and $\tau=\tau_k$, the Kraus coefficients in Eq.~(\ref{alphabeta}) for the states subject to Eq.~(\ref{condition}) can be approximated as
\begin{equation}
\begin{aligned}\label{alpham_approx}
\alpha_m(\tau_k)&=\cos\left[l_k\pi\sqrt{\frac{n_m+\nu/\mu }{n_t(k-1)+\nu/\mu}}\right]\\
&\approx\cos\left\{l_k\pi\left[1+\frac{1}{2}\frac{n_m-n_t(k-1)}{n_t(k-1)+\nu/\mu}\right]\right\}\\
&\approx(-1)^{l_k}\cos\left\{\frac{\pi[n_m-n_t(k-1)]}{2^k}\right\},
\end{aligned}
\end{equation}
and
\begin{equation}\label{betam_approx}
\beta_m(\tau_k)\approx\frac{\mathcal{Q}_m}{|\mathcal{Q}_m|}(-1)^{l_k}
\sin\left\{\frac{\pi[n_m-n_t(k-1)]}{2^k}\right\}.
\end{equation}
Then $\alpha_m(\tau_k)$ and $\beta_m(\tau_k)$ are unit in magnitude when $n_m-n_t(k-1)=2^k q$ and $n_m-n_t(k-1)=2^k(q+1/2)$ with $q\in\mathbb{Z}$, respectively. It means that the population on particular $|\lambda_m\rangle$'s could be conserved under measurements at the desired time points in Eq.~(\ref{tauk}) and the rest will be reduced. In addition, the approximation in Eqs.~(\ref{alpham_approx}) and (\ref{betam_approx}) will be more reasonable for a larger GPM label $n_t(k)$ or a larger average number of excitations with respect to the initial state of main system.

By Eq.~(\ref{alphabeta}), a nonvanishing detuning $\Delta$ yields $\delta_m<|\alpha_m|^2<1$ with $\delta_m\equiv\Delta^2/(4\Omega_m^2)$ and $0\leq|\beta_m|^2<1$. Then the discrepancy between the survive probabilities of the states in a generalized parity subspace relative to $n_t(k-1)$, i.e., $n_m-n_t(k-1)=2^k q$, and those of the other states will be reduced in comparison to the resonant case of $\Delta=0$. Nevertheless, $\delta_m\ll1$ provided that $|\Delta|\ll g|\mathcal{Q}_m|$. Then the correction induced by a nonvanishing $\Delta$ is perturbatively small and it can be rapidly suppressed to $\mathcal{O}(\delta_m^{N_{\alpha}})$ under repeated measurements, where $N_{\alpha}$ is the number of operations involving POVMs $V_{ee}(\tau)$ and $V_{gg}(\tau)$. Consequently, the influence of a small $\Delta$ on our protocol is ignorable~\cite{PRAZhang2026}.

\emph{Main result}. We find that if the GPM label is updated based on the following adaptive rule
\begin{equation}\label{adaptive_rule}
n_t(k+1)=n_t(k)+\left(1-\delta_{m_{k+1}m_{k}}\right)2^{k},\quad k\geq 0,
\end{equation}
where $m_k=e,g$ denotes the measurement outcome of the $k$th round and initially $m_0=e$ (a casual choice), then $k$ rounds of free joint-evolution and projection construct a generalized-parity measurement. More explicitly, if the $k$th outcome is $|e\rangle$, then the protocol realizes a ``diagonal'' GPM
\begin{equation}\label{MGPk}
\mathcal{M}_{\rm GP}^{(k)}\sim\sum_{n_m=n_t(k)\, {\rm mod}\,2^k}|\lambda_m\rangle\langle \lambda_m|.
\end{equation}
Otherwise, if the $k$th outcome is $|g\rangle$, the protocol realizes a ``displaced'' GPM
\begin{equation}\label{TGPk}
\mathcal{T}_{\rm GP}^{(k)}\sim\sum_{n_m=n_t(k)\, {\rm mod}\,2^k}|\lambda_{m+1}\rangle\langle \lambda_m|.
\end{equation}
Note ``$\sim$'' means the equivalence up to local phases that do not affect the population redistribution.

Equation~(\ref{adaptive_rule}) can be straightforwardly verified in the first round of evolution and measurement. The ancillary qubit is assumed in $|e\rangle$ from beginning. Then if the measurement outcome is again $|e\rangle$, the relevant measurement operator is given by
\begin{equation}
\begin{aligned}
V_{ee}(\tau_1)&\sim\sum_m\cos\left\{\frac{\pi[n_m-n_t(0)]}{2}\right\}
|\lambda_m\rangle\langle\lambda_m| \\
&\sim\sum_{n_m=n_t(0)\,{\rm mod}\,2}|\lambda_m\rangle\langle\lambda_m|
\end{aligned}
\end{equation}
due to Eqs.~(\ref{Vee}) and (\ref{alpham_approx}). Since the measurement outcome satisfies $m_1=m_0$, Eq.~(\ref{adaptive_rule}) yields an invariant label after the first round, e.g., $n_t(1)=n_t(0)$. In this case, we have a parity measurement in the normal sense,
\begin{equation}\label{Vee1}
V_{ee}(\tau_1)\sim\sum_{n_m=n_t(1)\,{\rm mod}\,2}
|\lambda_m\rangle\langle\lambda_m|=\mathcal{M}_{\rm GP}^{(1)}.
\end{equation}

Otherwise if the first-round measurement outcome is $|g\rangle$, the relevant measurement operator is obtained by Eqs.~(\ref{Vge}) and (\ref{betam_approx}) as
\begin{equation}
\begin{aligned}
V_{ge}(\tau_1)&\sim\sum_m\sin\left\{\frac{\pi[n_m-n_t(0)]}{2}\right\}
|\lambda_{m+1}\rangle\langle\lambda_m| \\
&\sim\sum_{n_m=n_t(0)+1\,{\rm mod}\,2}|\lambda_{m+1}\rangle\langle\lambda_m|.
\end{aligned}
\end{equation}
It projects the target system onto the subspace of odd parity relative to $n_t(0)$ and moves the conserved populations from $|\lambda_m\rangle$ to $|\lambda_{m+1}\rangle$ in the same time. The GPM label after the first round changes as $n_t(1)=n_t(0)+1$ due to Eq.~(\ref{adaptive_rule}) and $m_1\neq m_0$. Thus, we have
\begin{equation}\label{Vge1}
V_{ge}(\tau_1)\sim\sum_{n_m=n_t(1)\,{\rm mod}\,2}|\lambda_{m+1}\rangle\langle\lambda_m|=\mathcal{T}_{\rm GP}^{(1)}.
\end{equation}

Equations~(\ref{Vee1}) and~(\ref{Vge1}) have confirmed that the adaptive rule~(\ref{adaptive_rule}) holds after the first round of measurement. One can verify that it also holds after the second round by considering four possible measurement branches. For the branches that do not change the measurement outcome, i.e., $e\to e$ and $g\to g$, the survived components $|\lambda_m\rangle$ in the first round are further separated in the eigenbasis space, because the filtering condition is refined from modulo $2$ to modulo $4$ with the invariant GPM label. For the branches with complementary measurement outcome, i.e., $e\to g$ and $g\to e$, the survived components are determined by the shifted label. The effective measurement operator therefore still takes the form of either $\mathcal{M}_{\rm GP}^{(2)}$ or $\mathcal{T}_{\rm GP}^{(2)}$ after two rounds, depending exclusively on the second measurement outcome. The above two branches constitute the initial step of mathematical induction. Assuming that the result holds after the $k$th round and considering the same branch structure with the adaptive rule, one can verify that our result holds in the $(k+1)$th round. The details of the full induction are provided in Appendix~\ref{Proof}.

\section{Macroscopic Fock state generation}\label{Fock}

\subsection{Realizing adaptive GPM in JC model}\label{ideal}

In this section, we show that the adaptive rule can be directly implemented in JC model for deterministically realizing an effective GPM and then generating a giant Fock state. The full Hamiltonian of JC model reads
\begin{equation}
H=\frac{\omega_q}{2}\sigma_z+\omega_ca^\dagger a+g\left(\sigma_-a^\dagger+\sigma_+a\right),
\end{equation}
where $\omega_q$ and $\omega_c$ are the energy splitting of the ancillary qubit and the frequency of the resonator, respectively, and $g$ is the coupling strength. $a^\dagger$ $(a)$ is the creation (annihilation) operator of the target resonator. In the rotating frame with respect to $U_I(t)=\exp[i\omega_c(\sigma_z/2+a^\dagger a)t]$, the full Hamiltonian is expressed in the similar form of Eq.~(\ref{H_total}),
\begin{equation}\label{JCHam}
H_I=\frac{\Delta}{2}\sigma_z+g\left(\sigma_-a^\dagger+\sigma_+a\right),
\end{equation}
where $\Delta=\omega_q-\omega_c$ denotes the detuning between qubit and resonator. The creation operator takes the ladder form in the Fock space:
\begin{equation}
a^\dagger=\sum_{m=0}^\infty\sqrt{m+1}|m+1\rangle\langle m|,
\end{equation}
that mimics $Q^\dagger$ in Eq.~(\ref{Qdagger}) and the Fock basis $|m\rangle$ corresponds to the eigenbasis $|\lambda_m\rangle$ of $\Omega$ in Eq.~(\ref{OmegaLambda}). Consequently, $|\mathcal{Q}_m|^2=m+1$, satisfying the general spectrum condition in Eq.~(\ref{Qk}) with $\mu=1$, $\nu=1$, and $n_m=m$.

Coherent state can be readily prepared in experiments. And recent studies connected coherent and Fock states under a large-photon-number limit in a proper mode representation~\cite{PRLDescamps2024,OpticaQDescamps2026}. We then choose the coherent state $|\alpha\rangle$ with $|\alpha|^2\approx n_t(0)\gg 1$ as the starting point for generating large Fock states. Note the population distribution of the coherent state in Fock space is centered around $|n_t(0)\rangle$ and the population on $|m\rangle$ is negligible when $|m-n_t(0)|\gg\sqrt{n_t(0)}$. One can verify that all the Fock components in the dominantly populated regime $|m-n_t(0)|<\sqrt{n_t(0)}$ satisfy Eq.~(\ref{condition}). Using the evolution duration in Eq.~(\ref{tauk}) and the updating rule for the label in Eq.~(\ref{adaptive_rule}), a generalized parity measurement centered around the Fock component $|n_t(N)\rangle$ can be constructed after $N$ rounds of evolution and measurement:
\begin{equation}\label{MGP_JC}
\mathcal{M}_{\rm GP}^{(N)}=\sum_j |n_t(N)+2^N j\rangle\langle n_t(N)+2^N j|,
\end{equation}
if the measurement outcome in the last round is $|e\rangle$. Otherwise, we have
\begin{equation}\label{TGP_JC}
\mathcal{T}_{\rm GP}^{(N)}=\sum_j |n_t(N)+1+2^N j\rangle\langle n_t(N)+2^N j|,
\end{equation}
which is equivalent to performing a GPM of module $2^N$ centered around $|n_t(N)\rangle$ and then adding a photon to each Fock component.

\begin{figure}[htbp]
\centering
\includegraphics[width=0.9\linewidth]{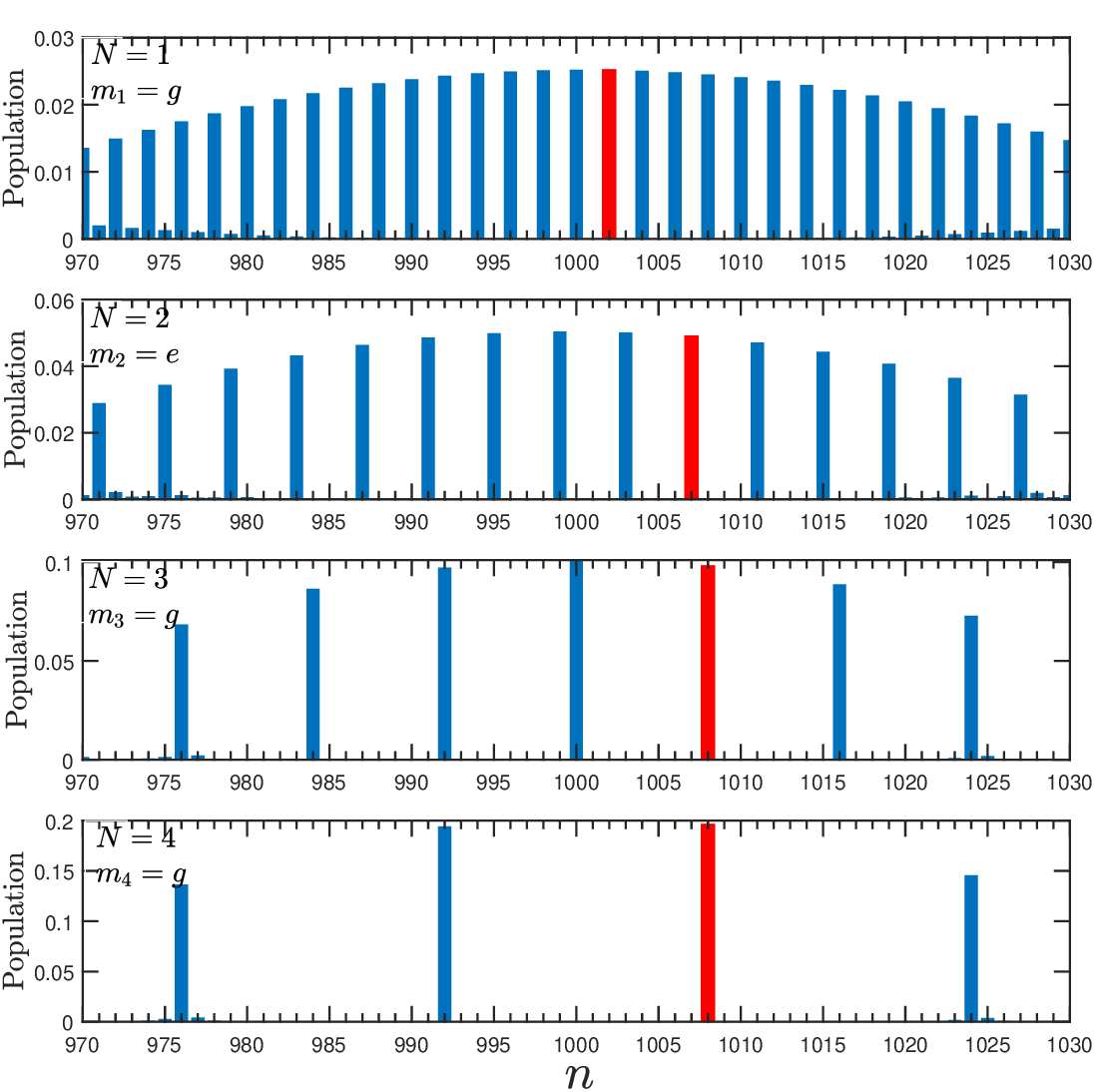}
\caption{An example about the variation of the population distribution in Fock space after the first $4$ rounds for a specific sequence of measurement results $\{|g\rangle, |e\rangle, |g\rangle, |g\rangle\}$. The initial GPM label is set as $n_t(0)=1000$. The red bar in each panel marks the updated GPM label $n_t(N)$ for measurements' outcome $m_N=e$ or $n_t(N)+1$ for $m_N=g$ due to Eq.~(\ref{adaptive_rule}).}\label{popu_dis}
\end{figure}

Figure~\ref{popu_dis} demonstrates an example of the population redistribution on the Fock bases after the first four rounds with a particular measurement-outcome sequence $\{|g\rangle, |e\rangle, |g\rangle, |g\rangle\}$. The main system starts from a coherent state $|\alpha\rangle$ with $|\alpha|^2=n_t(0)=1000$. The red bar in each panel marks the GPM label $n_t(N)$ for measurements' outcome $m_N=e$ or $n_t(N)+1$ for $m_N=g$, according to Eq.~(\ref{adaptive_rule}). Initially, the resonator is mainly populated around $|n_t(0)\rangle$ by a Gaussian shape. The population distribution is stepwisely concentrated to the generalized-parity subspace that formulates a structure symmetric to $|n_t(N)\rangle$ or $|n_t(N)+1\rangle$ as the protocol proceeds. This behavior is completely determined by GPMs $\mathcal{M}_{\rm GP}^{(N)}$ and $\mathcal{T}_{\rm GP}^{(N)}$ in Eqs.~(\ref{MGP_JC}) and (\ref{TGP_JC}), respectively. After $N$ rounds of evolution and measurement, every neighboring pair of the survived Fock components of a general parity are separated by $2^N$ in the Fock index. A high-fidelity Fock state thus emerges when only one of these survived components falls into the initially significantly populated regime, while the others are out of the regime, i.e., $2^N\gg\sqrt{n_t(0)}$ and then negligibly populated. Our protocol will ever continue in the presence of any measurement outcome $|e\rangle$ or $|g\rangle$. In other words, we take all the measurement trajectories into account rather than performing postselection at the end of each round. The population and fidelity are obtained by ensemble average over all possible sequences or measurement trajectories, unless otherwise stated.

The discrepancy between the last GPM label and the initial one, i.e., $\Delta n_t=n_t(N)-n_t(0)$, ranges from $0$ to $\sum_{k=0}^{N-1}2^k\sim 2^N$. For a tiny fraction of the measurement trajectories, $\Delta n_t$ might exceed the initial width of the significantly populated regime, which is of order $\sqrt{n_t(0)}$. In particular, when $\Delta n_t<2^N-\sqrt{n_t(0)}$, all the Fock components survived under $N$ rounds of evolution-measurement are shifted to the regime of negligible populations. These trajectories will reduce the success probability of producing a high-fidelity Fock state in terms of ensemble average, however, that can be alleviated for a target of an even larger number of excitations.

Given a sufficiently large number of average excitations, our protocol can generate a Fock state $|m\gg1\rangle$ with a high fidelity on average, as long as $n_t(0)\gg1$ and $|m-n_t(0)|<\sqrt{n_t(0)}$, although we cannot exactly know $m$ in advance. Since every evolution-measurement round doubles the distance between the survived Fock components, the measurement number required to distinguish a single Fock component roughly follows a logarithmic scaling law
\begin{equation}\label{scaling}
N\propto\log_2\sqrt{n_t(0)},
\end{equation}
the same as those reported in the previous GPM-based protocols for eigenstate preparation~\cite{NatPhysDeng2024,PRAZhang2026}. By Eqs.~(\ref{tauk}) and (\ref{Rabi_fre}), the measurement intervals roughly scale as $\tau_k\propto\sqrt{n_t(0)}$ for $n_t(0)\gg1$. The total running time of our protocol can then be estimated as $T\propto\sqrt{n_t(0)}\log_2\sqrt{n_t(0)}$.

For a given measurement record or trajectory $\mathbf{m}_N=(m_1,m_2,\cdots,m_N)$ with $m_k=e,g$, the effective Kraus operator turns out to be
\begin{equation}
K_{\mathbf{m}_N}=V_{m_Nm_{N-1}}(\tau_N)\cdots V_{m_2m_1}(\tau_2)V_{m_1m_0}(\tau_1),
\end{equation}
where $m_0=e$ indicates the initial state of the qubit. The relevant conditional state and the probability of this trajectory are given by
\begin{equation}
\rho_{\mathbf{m}_N}^{(N)}=\frac{K_{\mathbf{m}_N}\rho_cK_{\mathbf{m}_N}^\dagger}
{P_{\mathbf{m}_N}},\quad 
P_{\mathbf{m}_N}={\rm Tr}\left[K_{\mathbf{m}_N}\rho_cK_{\mathbf{m}_N}^\dagger\right],
\end{equation}
respectively. Despite the final distinguished Fock component is trajectory dependent, it is located in a limited regime around $|n_t(0)\rangle$. We therefore define the fidelity of the generated Fock state as
\begin{equation}
\mathcal{F}_{\mathbf{m}_N}=\max_m \langle m|\rho_{\mathbf{m}_N}^{(N)}|m\rangle, \quad |m-n_t(0)|<\sqrt{n_t(0)}.
\end{equation}
The fidelity in the following figures can then be evaluated by ensemble average:
\begin{equation}
\bar{\mathcal{F}}_N=\sum_{m_N=e,g}\cdots\sum_{m_1=e,g}P_{\mathbf{m}_N}\mathcal{F}_{\mathbf{m}_N}.
\end{equation}

\begin{figure}[htbp]
\centering
\includegraphics[width=0.9\linewidth]{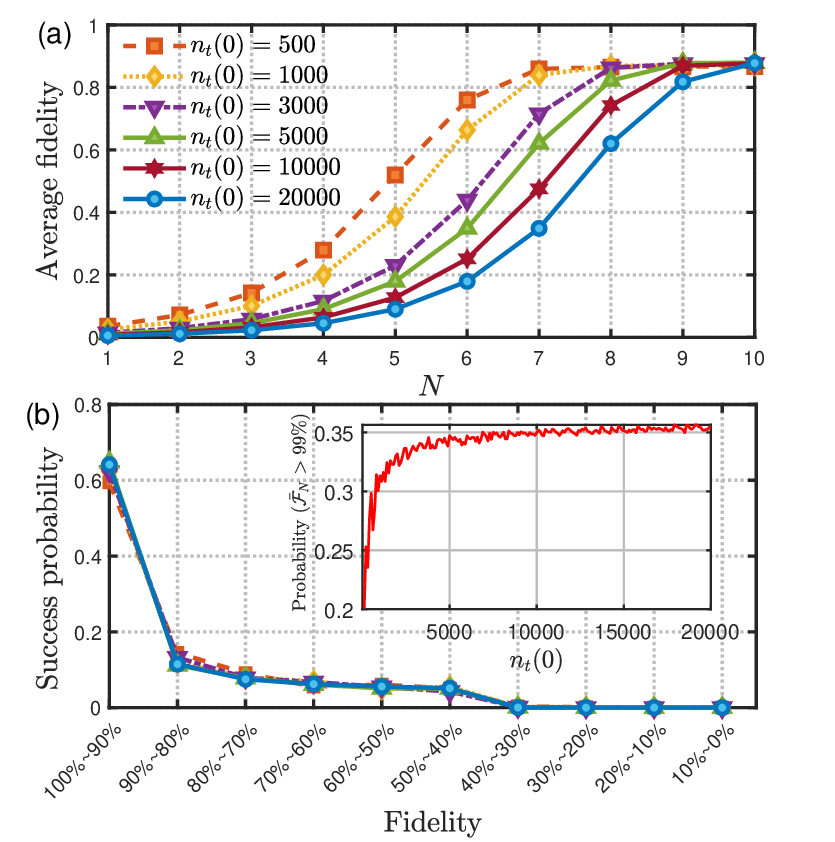}
\caption{(a) Average fidelity over all possible measurement trajectories and (b) probability distribution about the final fidelity over all the equiv-distance intervals for various initial coherent states $|\alpha=\sqrt{n_t(0)}\rangle$. Inset of (b) shows the probability about the final fidelity exceeding $99\%$ as a function of $n_t(0)$.}\label{Fidelity}
\end{figure}

Figure~\ref{Fidelity} demonstrates the performance of our adaptive protocol for the generation of large Fock-state from various coherent states $|\alpha=\sqrt{n_t(0)}\rangle$. The numerical simulation is performed over $500\leq n_t(0)\leq 20000$. This range is mainly constrained by the computer capacity (we only use a desktop with an Intel Core i7-7700 processor with 3.60 gigahertz in frequency and 16 gigabyte in memory) not by the protocol itself, which is irrespective of the excitation number of the target Fock state.

As shown in Fig.~\ref{Fidelity}(a), the fidelity averaged over all measurement trajectories can rapidly converge to about $87\%$ in less than $10$ rounds of measurement as expected by the logarithmic scaling behavior in Eq.~(\ref{scaling}). In Fig.~\ref{Fidelity}(b), we plot the success probability distribution of the final fidelity over the intervals spaced by $10\%$. It is found that the distribution is dominantly populated in the highest-fidelity interval and the probabilities of lower fidelities rapidly decline. We have $80\%$ measurement trajectories to produce a Fock state in the range of  $|20000-\sqrt{20000}\leq n_t\leq20000+\sqrt{20000}\rangle$ of a fidelity $\geq70\%$. The inset of Fig.~\ref{Fidelity}(b) shows the success probability that the final fidelity exceeds $99\%$ as a function of $n_t(0)$. The success probability of a near-unit-fidelity Fock state generation increases with $n_t(0)$ and then saturates in the large-$n_t(0)$ regime. It approaches $35\%$ when $n_t(0)\geq5000$, much higher than the protocols based on postselection, which is just the initial population on the target state~\cite{NatPhysDeng2024,PRAZhang2026}.

\subsection{Effect of readout error}\label{error}

In practice, the qubit readout in measurement can no longer remain faithful under detection errors, where the actual outcome $|e\rangle$ ($|g\rangle$) can be recorded as $g$ ($e$) with a finite error rate $\epsilon$. Such an error is especially relevant to the present adaptive protocol, because the measurement record determines the update of GPM and the duration of the next joint-evolution interval. An incorrect record can cause the subsequent adaptive setting inconsistent with the generalized-parity subspace associated with GPM for the correct measurement outcome.

\begin{figure}[htbp]
\centering
\includegraphics[width=0.9\linewidth]{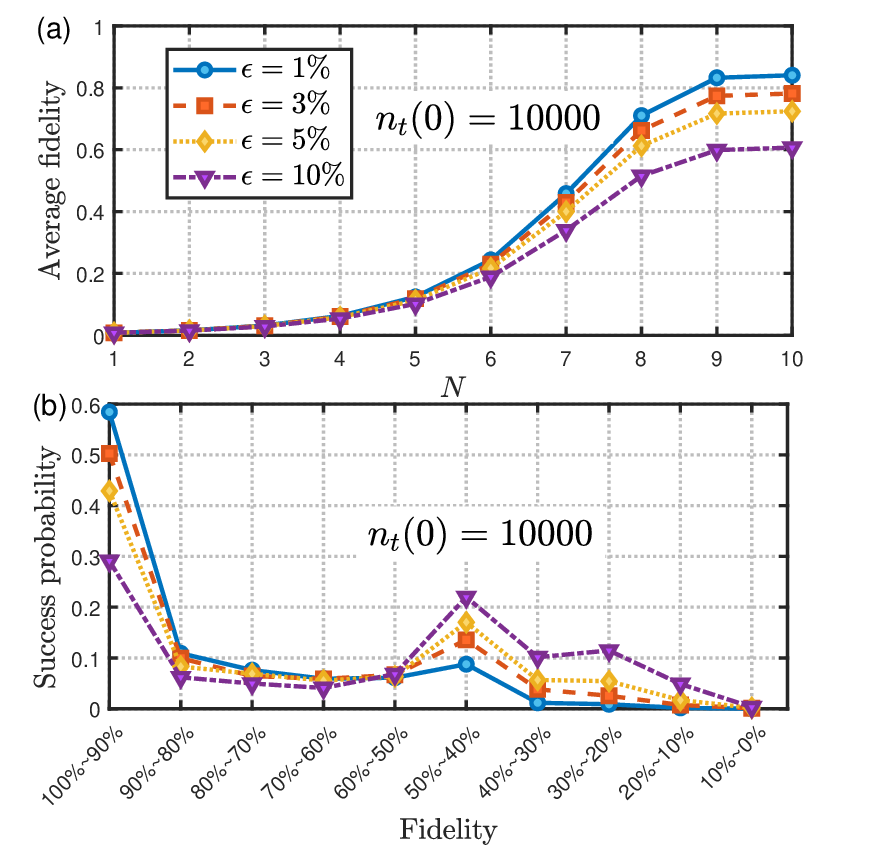}
\caption{Performance of imperfect adaptive GPM protocol. (a) Average fidelity as a function of the measurement number $N$ with various readout-error rates $\epsilon$. (b) Success probability distribution of the final fidelity after $N=10$ rounds over various intervals. }\label{measure_error}
\end{figure}

Figure~\ref{measure_error} presents the effect of measurement readout errors on the performance of our adaptive GPM protocol. The duration for each interval between measurements is still updated by the rule in Eq.~(\ref{adaptive_rule}) for GPM label and the formula in Eq.~(\ref{tauk}). However, there is a probability of $\epsilon$ by which the correct measurement outcome will incorrectly flip to its complement, i.e., $|e\rangle\rightarrow|g\rangle$ or $|g\rangle\rightarrow|e\rangle$. The numerical results become convergent under about $10^4$ measurement trajectories by Monte Carlo simulation. As shown in Fig.~\ref{measure_error}(a), the average fidelity still increases monotonically with the measurement number, although it is gradually suppressed by increasing readout-error rate. With $\epsilon=1\%$, $3\%$, $5\%$, and $10\%$, the average fidelity approaches $84\%$, $78\%$, $72\%$, and $60\%$, respectively. A more detailed picture about this degradation in performance can be found in Fig.~\ref{measure_error}(b), which presents the success probability distribution about the final fidelity. For a low rate of readout error with $\epsilon<5\%$, the probability for the highest-fidelity interval $90\%-100\%$ still dominates the whole distribution, showing that a substantial fraction of measurement trajectories can generate a giant Fock state with a near-unit fidelity. With increasing $\epsilon$, the probability weight gradually shifts from the higher-fidelity interval to lower-fidelity intervals. In particular, the probability for obtaining a Fock state with fidelity lower than $50\%$ is about $48\%$ when $\epsilon=10\%$. Since the readout-error rates in the state-of-art circuit-QED platforms are typically not greater than $3\%$~\cite{PRLJeffrey2014,PRAppliedWalter2017,PRAppliedHeinsoo2018,PRXQMarxer2026}, these numerical results indicate that our adaptive GPM mechanism is compatible with the readout accuracy available in experiments.

\subsection{Robustness against thermal mixture}\label{thermal}

Our protocol for Fock-state generation combines the efficiency of GPM following a logarithmic scaling law and the compatibility with the system's initial state. It is essentially an adaptive filter for unwanted Fock components through rounds of joint evolution and projective measurement on qubit, and therefore applies to both pure and mixed states. A realistic cavity- or circuit-QED system or trapped ion cannot be completely isolated from a finite-temperature environment, that brings the resonator into a thermal mixture before it is prepared as a coherent state by a displacement operation $D(\alpha)=\exp(\alpha a^\dagger-\alpha^* a)$. Thus we consider a displaced thermal state $D(\alpha)\rho_{\rm th}D^\dagger(\alpha)$, given that the photon population is still dominantly concentrated about $|n_t(0)\rangle$. Here $\alpha=\sqrt{n_t(0)}$ and $\rho_{\rm th}=e^{-\beta\omega_ca^\dagger a}/{\rm Tr}[e^{-\beta \omega_ca^\dagger a}]$ is the initial thermal state of the target resonator.

\begin{figure}[htbp]
\centering
\includegraphics[width=0.9\linewidth]{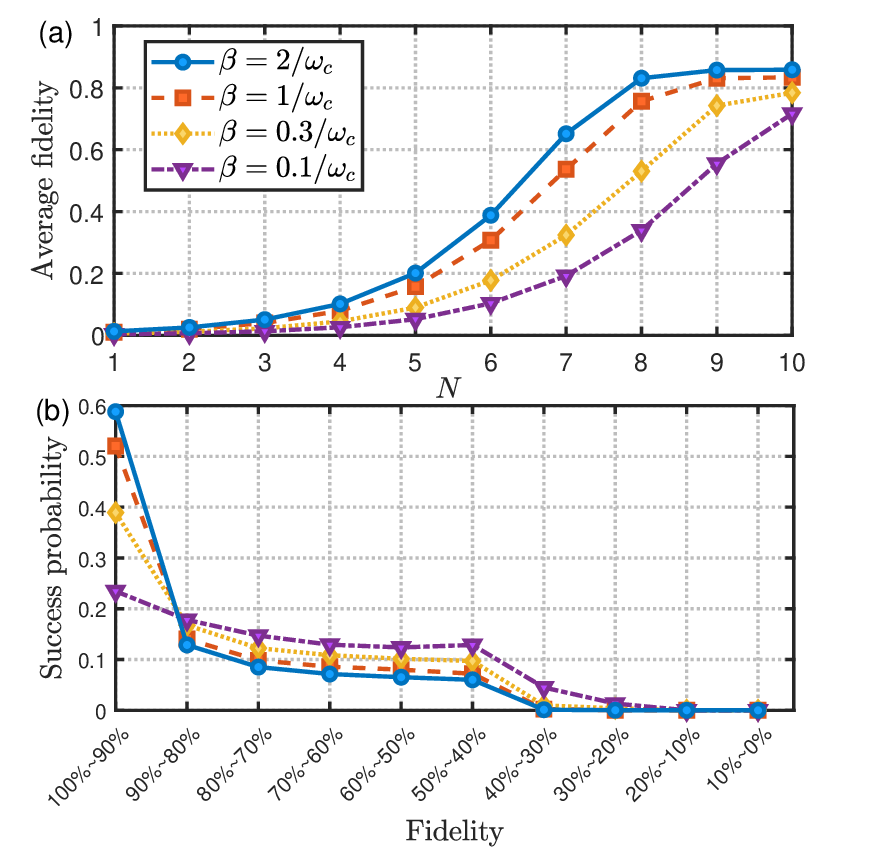}
\caption{Performance of our protocol when the resonator is prepared as a displaced thermal state $D(\alpha)\rho_{\rm th}D^\dagger(\alpha)$ under various inverse temperatures $\beta$. (a) Average fidelity as a function of the measurement number $N$. (b) Success probability distribution of the final fidelity. $n_t(0)=3000$.}\label{Thermal_Fedility}
\end{figure}

Figure~\ref{Thermal_Fedility} shows the performance of our protocol for displaced thermal states with various inverse temperatures $\beta$. On one hand, the adaptive GPM remains effective under a certain degree of classical mixture. As in Fig.~\ref{Thermal_Fedility}(a), the average fidelity still increases with the measurement number $N$ for all $\beta$. On the other hand, the generation efficiency gradually decreases and the final fidelity reduces with increasing temperature or decreasing $\beta$, since a smaller $\beta$ means a broader Fock-state distribution. After $10$ measurements, the average fidelity approaches $85\%$, $83\%$, $78\%$, and $71\%$ for $\beta=2/\omega_c$, $1/\omega_c$, $0.3/\omega_c$, and $0.1/\omega_c$, respectively. In Fig.~\ref{Thermal_Fedility}(b) for the probability distribution of the final fidelity, one can find that the curve is flattened as the temperature is enhanced. Nevertheless, we have $79\%$, $74\%$, $67\%$, and $56\%$ in success probability to generate a Fock state about $|3000-\sqrt{3000}\leq n_t\leq3000+\sqrt{3000}\rangle$ of a fidelity $\geq70\%$, for $\beta=2/\omega_c$, $1/\omega_c$, $0.3/\omega_c$, and $0.1/\omega_c$, respectively. Our protocol is thus robust against moderate temperatures.

\begin{figure}[htbp]
\centering
\includegraphics[width=0.95\linewidth]{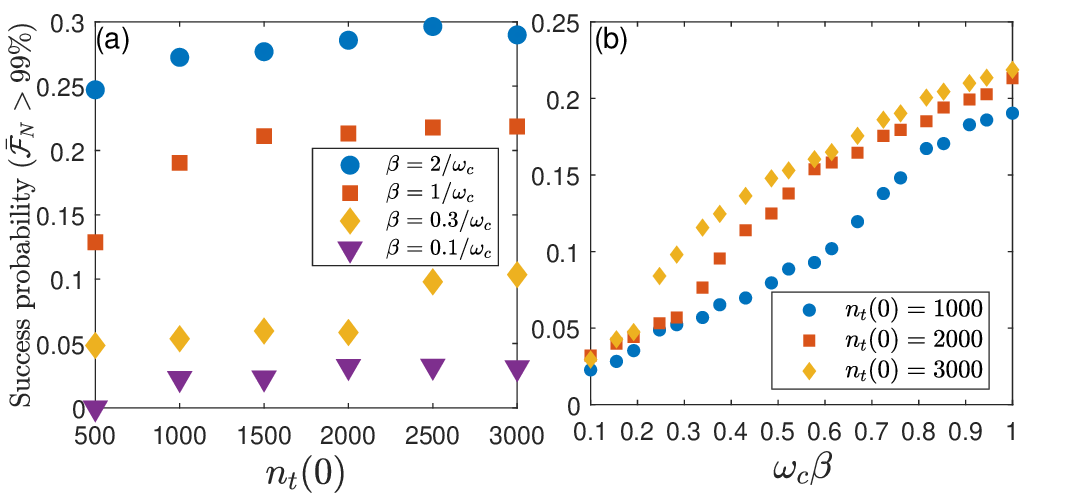}
\caption{Success probability for obtaining a Fock state by $N=10$ rounds of evolution and measurement with a fidelity higher than $99\%$ as a functions of (a) $n_t(0)$ for various $\beta$'s and of (b) the inverse temperature $\beta$ for various $n_t(0)$'s. The initial state is the displaced thermal state $D(\alpha)\rho_{\rm th}D^\dagger(\alpha)$.}\label{Thermal_Probability}
\end{figure}

Figure~\ref{Thermal_Probability}(a) plots the success probability that the average fidelity $\geq99\%$ under $N=10$ rounds of measurement as a function of the initial GPM label $n_t(0)$ for displaced thermal states with various temperatures. The probability roughly increases with $n_t(0)$ for each fixed $\beta$. It still reflects the characteristic of our protocol implied in Eq.~(\ref{alpham_approx}) about the Kraus coefficient, that the protocol is more appropriate for generating larger Fock states. With $\beta=2/\omega_c$, $1/\omega_c$, $0.3/\omega_c$, and $0.1/\omega_c$, the probability of obtaining a Fock state of $\sim3000$ excitations with a fidelity above $99\%$ is $29\%$, $22\%$, $10\%$, and $3\%$, respectively. The temperature effect can be also shown in Fig.~\ref{Thermal_Probability}(b) by the near-unit-fidelity probability as a function of $\beta$ for $n_t(0)=1000$, $2000$, and $3000$. The probability increases monotonically with both $\beta$ and $n_t(0)$. This behavior is consistent with the wider population distribution of the displaced thermal state at a higher temperature, which reduces the filtering efficiency of GPM. Nevertheless, it demonstrates both robustness and scalability of our protocol free of the pure-state initialization.

\section{Conclusion}\label{conclusion}

In summary, we have presented a general protocol for quantum state engineering based on adaptive measurement without postselection. By repeated projective measurements on an ancillary qubit coupled to a high-dimensional target system of discrete spectrum through an exchange-type interaction, a GPM up to local phases can be deterministically constructed with the measurement intervals adaptively updated in response to the last measurement outcome. For JC model, one of the fundamental interactions of quantum optics and realized across platforms, numerical simulations show that a giant Fock state with photon numbers of about 20000 can be generated within $10$ measurement rounds. And the success probabilities of the final fidelities above $99\%$ and $90\%$ approach $35\%$ and $65\%$, respectively, which are nearly two orders in magnitude greater than that of ideal postselection for the same initial coherent state $|\alpha=\mathcal{O}(10^2)\rangle$.

Our protocol is free of adaptive driving fields, gate operations, changes of measurement basis, and nonlinear interactions. The only adaptive control is over the duration of the joint free evolution of the qubit-resonator system. In practice, our protocol is robust against the readout-error on measurement outcome. The average fidelity about generating a Fock state of $|n_t\approx10000\rangle$ remains above $78\%$ in the presence of an error rate $\epsilon=3\%$. It is also fault-tolerant with detuning deviation and thermal initialization. Even if the resonator starts from a displaced thermal state at $\beta=0.1/\omega_c$, a Fock state of $|n_t\approx3000\rangle$ can be generated with an average fidelity above $71\%$ by $10$ rounds of measurements. It conforms that our adaptive GPM protocol provides a convenient and experimentally accessible route to macroscopic nonclassical-state preparation.

\section*{Acknowledgments}

We acknowledge grant support from the National Natural Science Foundation of China (Grant No. U25A20199) and the ``Pioneer'' and ``Leading Goose'' R\&D of Zhejiang Province (Grant No. 2025C01028).

\appendix

\section{Proof of the adaptive rule}\label{Proof}

This appendix presents the proof of the adaptive rule (\ref{adaptive_rule}) by mathematical induction. Actually the first step of the induction has been presented in Sec.~\ref{adap_rule} to yield either Eq.~(\ref{Vee1}) or Eq.~(\ref{Vge1}) under two different branches.

We first consider the branch that the outcome of the first evolution-measurement round is $|e\rangle$. By Eq.~(\ref{Vee1}), the resulting operator is the parity-measurement operator $\mathcal{M}_{\rm GP}^{(1)}$. Then in the second round, if the outcome is again $|e\rangle$, the resulting operator is
\begin{equation}
\begin{aligned}
&V_{ee}(\tau_2)V_{ee}(\tau_1)  \\
&\sim\sum_m\cos\left\{\frac{\pi[n_m-n_t(1)]}{4}\right\}|\lambda_m\rangle\langle\lambda_m|\\
&\times\sum_{n_{m'}= n_t(1)\,{\rm mod}\,2}|\lambda_{m'}\rangle\langle\lambda_{m'}|
\sim\sum_{n_m=n_t(1)\,{\rm mod}\,4}|\lambda_m\rangle\langle\lambda_m|
\end{aligned}
\end{equation}
by Eqs.~(\ref{Vee}) and (\ref{alpham_approx}). Since the measurement outcomes satisfy $m_2=m_1$, we have $n_t(2)=n_t(1)$. The resulting operator can then be formally expressed by the GPM operator with parity $2^2$:
\begin{equation}
V_{ee}(\tau_2)V_{ee}(\tau_1)\sim\sum_{n_m= n_t(2)\,{\rm mod}\,4}|\lambda_m\rangle\langle\lambda_m|=\mathcal{M}_{\rm GP}^{(2)} .
\end{equation}

However, if the first-round outcome is $|e\rangle$ and the second-round outcome is $|g\rangle$, then by Eqs.~(\ref{Vge}) and (\ref{betam_approx}), the product of the two Kraus operators is
\begin{equation}
\begin{aligned}
&V_{ge}(\tau_2)V_{ee}(\tau_1)  \\
&\sim\sum_m\sin\left\{\frac{\pi[n_m-n_t(1)]}{4}\right\}|\lambda_{m+1}\rangle\langle\lambda_m|\\
&\times\sum_{n_{m'}=n_t(1)\,{\rm mod}\,2}|\lambda_{m'}\rangle\langle\lambda_{m'}| \\
&\sim\sum_{n_m=n_t(1)+2\,{\rm mod}\,4}|\lambda_{m+1}\rangle\langle\lambda_m| .
\end{aligned}
\end{equation}
In this case, since $m_2\neq m_1$, Eq.~(\ref{adaptive_rule}) renders $n_t(2)=n_t(1)+2^1$. Hence, we have a displaced GPM operator:
\begin{equation}
V_{ge}(\tau_2)V_{ee}(\tau_1)\sim\sum_{n_m= n_t(2)\,{\rm mod}\,4}|\lambda_{m+1}\rangle\langle\lambda_m|=\mathcal{T}_{\rm GP}^{(2)}.
\end{equation}

Next we consider the branch in which the first-round measurement outcome is $|g\rangle$. From Eq.~(\ref{Vge1}), the relevant measurement operator is $\mathcal{T}_{\rm GP}^{(1)}$. If the second-round outcome is again $|g\rangle$, the Kraus operator induced by the second round of evolution and measurement is $V_{gg}(\tau_2)$. From Eqs.~(\ref{Vgg}) and (\ref{alpham_approx}), the measurement operator after two rounds reads
\begin{equation}
\begin{aligned}
&V_{gg}(\tau_2)V_{ge}(\tau_1) \\
&\sim\sum_m\cos\left\{\frac{\pi[n_{m-1}-n_t(1)]}{4}\right\}|\lambda_m\rangle\langle\lambda_m|\\
&\times\sum_{n_{m'}=n_t(1)\,{\rm mod}\,2}|\lambda_{m'+1}\rangle\langle\lambda_{m'}| \\
&\sim\sum_{n_m=n_t(1)\,{\rm mod}\,4}|\lambda_{m+1}\rangle\langle\lambda_m|.
\end{aligned}
\end{equation}
Since $m_2= m_1$, we have $n_t(2)=n_t(1)$. Consequently,
\begin{equation}
V_{gg}(\tau_2)V_{ge}(\tau_1)\sim\sum_{n_m= n_t(2)\,({\rm mod}\,4)}|\lambda_{m+1}\rangle\langle\lambda_m|=\mathcal{T}_{\rm GP}^{(2)} .
\end{equation}

Lastly, if the first-round outcome is $|g\rangle$ and the second-round outcome
is $|e\rangle$, the Kraus operator of the second round is then $V_{eg}(\tau_2)$. In this case, the measurement operator after two rounds reads
\begin{equation}
\begin{aligned}
&V_{eg}(\tau_2)V_{ge}(\tau_1) \\
&\sim\sum_m\sin\left\{\frac{\pi[n_{m-1}-n_t(1)]}{4}\right\}|\lambda_{m-1}\rangle\langle\lambda_m|\\
&\times \sum_{n_{m'}= n_t(1)\,{\rm mod}\,2}|\lambda_{m'+1}\rangle\langle\lambda_{m'}| \\
&\sim\sum_{n_m=n_t(1)+2\,{\rm mod}\,4}|\lambda_m\rangle\langle\lambda_m|.
\end{aligned}
\end{equation}
Since $m_2\neq m_1$, we have $n_t(2)=n_t(1)+2^1$. Thus,
\begin{equation}
V_{eg}(\tau_2)V_{ge}(\tau_1)
\sim\sum_{n_m= n_t(2)\,{\rm mod}\,4}|\lambda_m\rangle\langle\lambda_m|=\mathcal{M}_{\rm GP}^{(2)}.
\end{equation}
We now complete the second step for the proof by mathematical induction for $N=2$ rounds of evolution and measurement, given all the four possible branches have been explicitly confirmed. Assume that the effective measurement operator after $k$ rounds follows that in Eq.~(\ref{MGPk}) or Eq.~(\ref{TGPk}) when the measurement outcome of the $k$th round is $|e\rangle$ or $|g\rangle$, respectively. We can show that this claim still holds after the $(k+1)$th round.

First, suppose that the $k$th outcome is $|e\rangle$, i.e., $m_k=e$. If the $(k+1)$th outcome is also $|e\rangle$, the relevant Kraus operator is $V_{ee}(\tau_{k+1})$. By Eqs.~(\ref{Vee}), (\ref{alpham_approx}), and (\ref{MGPk}), we have
\begin{equation}
\begin{aligned}
&V_{ee}(\tau_{k+1})\mathcal{M}_{\rm GP}^{(k)} \\
&\sim\sum_m\cos\left\{\frac{\pi[n_m-n_t(k)]}{2^{k+1}}\right\}|\lambda_m\rangle\langle\lambda_m|\\
&\times\sum_{n_{m'}= n_t(k)\,{\rm mod}\,2^k}|\lambda_{m'}\rangle\langle\lambda_{m'}| \\
&\sim\sum_{n_m=n_t(k)\,{\rm mod}\,2^{k+1}}|\lambda_m\rangle\langle\lambda_m|.
\end{aligned}
\end{equation}
Since $m_{k+1}=m_k$, the adaptive rule~(\ref{adaptive_rule}) renders $n_t(k+1)=n_t(k)$. Consequently,
\begin{equation}
\begin{aligned}
&V_{ee}(\tau_{k+1})\mathcal{M}_{\rm GP}^{(k)}\\
&\sim\sum_{n_m= n_t(k+1)\,{\rm mod}\,2^{k+1}}|\lambda_m\rangle\langle\lambda_m|=\mathcal{M}_{\rm GP}^{(k+1)}.
\end{aligned}
\end{equation}

If the $k$th outcome is $|e\rangle$ yet the $(k+1)$th outcome is $|g\rangle$, the relevant Kraus operator is $V_{ge}(\tau_{k+1})$. Acting on Eq.~(\ref{MGPk}), we find
\begin{equation}
\begin{aligned}
&V_{ge}(\tau_{k+1})\mathcal{M}_{\rm GP}^{(k)} \\
&\sim\sum_m\sin\left\{\frac{\pi[n_m-n_t(k)]}{2^{k+1}}\right\}|\lambda_{m+1}\rangle\langle\lambda_m|\\
&\times\sum_{n_{m'}=n_t(k)\,{\rm mod}\,2^k}|\lambda_{m'}\rangle\langle\lambda_{m'}| \\
&\sim\sum_{n_m=n_t(k)+2^k\,{\rm mod}\, 2^{k+1}}|\lambda_{m+1}\rangle\langle\lambda_m| .
\end{aligned}
\end{equation}
Since the qubit flips from $|e\rangle$ to $|g\rangle$, the adaptive rule gives rise to $n_t(k+1)=n_t(k)+2^k$. Hence,
\begin{equation}
\begin{aligned}
&V_{ge}(\tau_{k+1})\mathcal{M}_{\rm GP}^{(k)}\\
&\sim\sum_{n_m= n_t(k+1)\,{\rm mod}\,2^{k+1}}|\lambda_{m+1}\rangle\langle\lambda_m|=\mathcal{T}_{\rm GP}^{(k+1)}.
\end{aligned}
\end{equation}

Next, suppose that the $k$th outcome is $|g\rangle$. Then the qubit is in
$|g\rangle$ at the beginning of the $(k+1)$th round. If the $(k+1)$th outcome is
also $|g\rangle$, the relevant Kraus operator is $V_{gg}(\tau_{k+1})$. Acting on Eq.~(\ref{TGPk}), we have
\begin{equation}
\begin{aligned}
&V_{gg}(\tau_{k+1})\mathcal{T}_{\rm GP}^{(k)} \\
&\sim\sum_m\cos\left\{\frac{\pi[n_{m-1}-n_t(k)]}{2^{k+1}}\right\}|\lambda_m\rangle\langle\lambda_m|\\
&\times\sum_{n_{m'}= n_t(k)\,{\rm mod}\,2^k}|\lambda_{{m'}+1}\rangle\langle\lambda_{m'}|\\
&\sim\sum_{n_m=n_t(k)\,{\rm mod}\,2^{k+1}}|\lambda_{m+1}\rangle\langle\lambda_m|.
\end{aligned}
\end{equation}
Since $m_{k+1}=m_k$, we have $n_t(k+1)=n_t(k)$. Then we have
\begin{equation}
\begin{aligned}
 &V_{gg}(\tau_{k+1})\mathcal{T}_{\rm GP}^{(k)}\\
 &\sim\sum_{n_m=n_t(k+1)\,{\rm mod}\,2^{k+1}}|\lambda_{m+1}\rangle\langle\lambda_m|=\mathcal{T}_{\rm GP}^{(k+1)} .
\end{aligned}
\end{equation}

Finally, if the $k$th outcome is $|g\rangle$ yet the $(k+1)$th outcome is $|e\rangle$, the relevant Kraus operator is $V_{eg}(\tau_{k+1})$. Acting on Eq.~(\ref{TGPk}), we have
\begin{equation}
\begin{aligned}
&V_{eg}(\tau_{k+1})\mathcal{T}_{\rm GP}^{(k)} \\
&\sim\sum_m\sin\left\{\frac{\pi[n_{m-1}-n_t(k)]}{2^{k+1}}\right\}|\lambda_{m-1}\rangle\langle\lambda_m|\\
&\times \sum_{n_{m'}= n_t(k)\,{\rm mod}\,2^k}|\lambda_{{m'}+1}\rangle\langle\lambda_{m'}|\\
&\sim\sum_{n_m=n_t(k)+2^k\,{\rm mod}\,2^{k+1}}|\lambda_m\rangle\langle\lambda_m| .
\end{aligned}
\end{equation}
Since the qubit flips from $|g\rangle$ to $|e\rangle$, the adaptive rule yields $n_t(k+1)=n_t(k)+2^k$. Then we end up with
\begin{equation}
\begin{aligned}
&V_{eg}(\tau_{k+1})\mathcal{T}_{\rm GP}^{(k)}\\
&\sim\sum_{n_m= n_t(k+1)\,{\rm mod}\,2^{k+1}}|\lambda_m\rangle\langle\lambda_m|=\mathcal{M}_{\rm GP}^{(k+1)}.
\end{aligned}
\end{equation}

We therefore exhaust all the four possible branches in the $(k+1)$th round. The proof of induction is completed.

\bibliographystyle{apsrevlong}
\bibliography{ref}

@article{RMPBraunstein2005,
  title = {Quantum information with continuous variables},
  author = {Braunstein, Samuel L. and van Loock, Peter},
  journal = {Rev. Mod. Phys.},
  volume = {77},
  issue = {2},
  pages = {513--577},
  numpages = {0},
  year = {2005},
  month = {Jun},
  publisher = {American Physical Society},
  doi = {10.1103/RevModPhys.77.513},
  url = {https://link.aps.org/doi/10.1103/RevModPhys.77.513}
}

@article{RMPWeedbrook2012,
  title = {Gaussian quantum information},
  author = {Weedbrook, Christian and Pirandola, Stefano and Garc\'{\i}a-Patr\'on, Ra\'ul and Cerf, Nicolas J. and Ralph, Timothy C. and Shapiro, Jeffrey H. and Lloyd, Seth},
  journal = {Rev. Mod. Phys.},
  volume = {84},
  issue = {2},
  pages = {621--669},
  numpages = {0},
  year = {2012},
  month = {May},
  publisher = {American Physical Society},
  doi = {10.1103/RevModPhys.84.621},
  url = {https://link.aps.org/doi/10.1103/RevModPhys.84.621}
}

@article{PRLPeaudecerf2014,
  title = {Adaptive quantum nondemolition measurement of a photon number},
  author = {Peaudecerf, B. and Rybarczyk, T. and Gerlich, S. and Gleyzes, S. and Raimond, J. M. and Haroche, S. and Dotsenko, I. and Brune, M.},
  journal = {Phys. Rev. Lett.},
  volume = {112},
  issue = {8},
  pages = {080401},
  numpages = {5},
  year = {2014},
  month = {Feb},
  publisher = {American Physical Society},
  doi = {10.1103/PhysRevLett.112.080401},
  url = {https://link.aps.org/doi/10.1103/PhysRevLett.112.080401}
}

@article{NatrueGuerlin2007,
  author = {Guerlin, C. and Bernu, J. and Del{\'e}glise, S. and Sayrin, C.
            and Gleyzes, S. and Kuhr, S. and Brune, M. and Raimond, J.-M.
            and Haroche, S.},
  title = {Progressive field-state collapse and quantum non-demolition photon counting},
  journal = {Nature},
  volume = {448},
  pages = {889--893},
  year = {2007},
  doi = {10.1038/nature06057}
}

@article{NatrueDeleglise2008,
  author = {Del{\'e}glise, S. and Dotsenko, I. and Sayrin, C. and Bernu, J.
            and Brune, M. and Raimond, J.-M. and Haroche, S.},
  title = {Reconstruction of non-classical cavity field states with snapshots of their decoherence},
  journal = {Nature},
  volume = {455},
  pages = {510--514},
  year = {2008},
  doi = {10.1038/nature07288}
}

@article{RMPBlais2021,
  title = {Circuit quantum electrodynamics},
  author = {Blais, Alexandre and Grimsmo, Arne L. and Girvin, S. M. and Wallraff, Andreas},
  journal = {Rev. Mod. Phys.},
  volume = {93},
  issue = {2},
  pages = {025005},
  numpages = {72},
  year = {2021},
  month = {May},
  publisher = {American Physical Society},
  doi = {10.1103/RevModPhys.93.025005},
  url = {https://link.aps.org/doi/10.1103/RevModPhys.93.025005}
}

@article{PRLMeekhof1996,
  title = {Generation of nonclassical motional states of a trapped atom},
  author = {Meekhof, D. M. and Monroe, C. and King, B. E. and Itano, W. M. and Wineland, D. J.},
  journal = {Phys. Rev. Lett.},
  volume = {76},
  issue = {11},
  pages = {1796--1799},
  numpages = {0},
  year = {1996},
  month = {Mar},
  publisher = {American Physical Society},
  doi = {10.1103/PhysRevLett.76.1796},
  url = {https://link.aps.org/doi/10.1103/PhysRevLett.76.1796}
}

@article{PRLBenKish2003,
  title = {Experimental demonstration of a technique to generate arbitrary quantum superposition states of a harmonically bound spin-$1/2$ particle},
  author = {Ben-Kish, A. and DeMarco, B. and Meyer, V. and Rowe, M. and Britton, J. and Itano, W. M. and Jelenkovi\ifmmode \acute{c}\else \'{c}\fi{}, B. M. and Langer, C. and Leibfried, D. and Rosenband, T. and Wineland, D. J.},
  journal = {Phys. Rev. Lett.},
  volume = {90},
  issue = {3},
  pages = {037902},
  numpages = {4},
  year = {2003},
  month = {Jan},
  publisher = {American Physical Society},
  doi = {10.1103/PhysRevLett.90.037902},
  url = {https://link.aps.org/doi/10.1103/PhysRevLett.90.037902}
}

@article{PRLMatsos2024,
  title = {Robust and deterministic preparation of bosonic logical states in a trapped ion},
  author = {Matsos, V. G. and Valahu, C. H. and Navickas, T. and Rao, A. D. and Millican, M. J. and Kolesnikow, X. C. and Biercuk, M. J. and Tan, T. R.},
  journal = {Phys. Rev. Lett.},
  volume = {133},
  issue = {5},
  pages = {050602},
  numpages = {7},
  year = {2024},
  month = {Jul},
  publisher = {American Physical Society},
  doi = {10.1103/PhysRevLett.133.050602},
  url = {https://link.aps.org/doi/10.1103/PhysRevLett.133.050602}
}

@article{PRXQuantumValahu2024,
  title = {Benchmarking bosonic modes for quantum information with randomized displacements},
  author = {Valahu, Christophe H. and Navickas, Tomas and Biercuk, Michael J. and Tan, Ting Rei},
  journal = {PRX Quantum},
  volume = {5},
  issue = {4},
  pages = {040337},
  numpages = {12},
  year = {2024},
  month = {Dec},
  publisher = {American Physical Society},
  doi = {10.1103/PRXQuantum.5.040337},
  url = {https://link.aps.org/doi/10.1103/PRXQuantum.5.040337}
}

@article{RMPAspelmeyer2014,
  title = {Cavity optomechanics},
  author = {Aspelmeyer, Markus and Kippenberg, Tobias J. and Marquardt, Florian},
  journal = {Rev. Mod. Phys.},
  volume = {86},
  issue = {4},
  pages = {1391--1452},
  numpages = {62},
  year = {2014},
  month = {Dec},
  publisher = {American Physical Society},
  doi = {10.1103/RevModPhys.86.1391},
  url = {https://link.aps.org/doi/10.1103/RevModPhys.86.1391}
}

@article{PhysRepZare2022,
  author = {Zare Rameshti, Babak and Viola Kusminskiy, Silvia and Haigh, James A. and Usami, Koji and Lachance-Quirion, Dany and Nakamura, Yasunobu and Hu, Can-Ming and Tang, Hong X. and Bauer, Gerrit E. W. and Blanter, Yaroslav M.},
  title = {Cavity magnonics},
  journal = {Phys. Rep.},
  volume = {979},
  pages = {1--61},
  year = {2022},
  doi = {10.1016/j.physrep.2022.06.001}
}

@article{RMPCaves1994,
  title = {Quantum limits on bosonic communication rates},
  author = {Caves, Carlton M. and Drummond, P. D.},
  journal = {Rev. Mod. Phys.},
  volume = {66},
  issue = {2},
  pages = {481--537},
  numpages = {0},
  year = {1994},
  month = {Apr},
  publisher = {American Physical Society},
  doi = {10.1103/RevModPhys.66.481},
  url = {https://link.aps.org/doi/10.1103/RevModPhys.66.481}
}

@article{NatureBouwmeester1997,
  author = {Bouwmeester, Dik and Pan, Jian-Wei and Mattle, Klaus and Eibl, Manfred and Weinfurter, Harald and Zeilinger, Anton},
  title = {Experimental quantum teleportation},
  journal = {Nature},
  volume = {390},
  pages = {575--579},
  year = {1997},
  doi = {10.1038/37539}
}

@article{PRLLloyd1999,
  title = {Quantum Computation over Continuous Variables},
  author = {Lloyd, Seth and Braunstein, Samuel L.},
  journal = {Phys. Rev. Lett.},
  volume = {82},
  issue = {8},
  pages = {1784--1787},
  numpages = {0},
  year = {1999},
  month = {Feb},
  publisher = {American Physical Society},
  doi = {10.1103/PhysRevLett.82.1784},
  url = {https://link.aps.org/doi/10.1103/PhysRevLett.82.1784}
}

@article{PRXMichael2016,
  title = {New Class of Quantum Error-Correcting Codes for a Bosonic Mode},
  author = {Michael, Marios H. and Silveri, Matti and Brierley, R. T. and Albert, Victor V. and Salmilehto, Juha and Jiang, Liang and Girvin, S. M.},
  journal = {Phys. Rev. X},
  volume = {6},
  issue = {3},
  pages = {031006},
  numpages = {26},
  year = {2016},
  month = {Jul},
  publisher = {American Physical Society},
  doi = {10.1103/PhysRevX.6.031006},
  url = {https://link.aps.org/doi/10.1103/PhysRevX.6.031006}
}

@article{PRLHeeres2015,
  title = {Cavity state manipulation using photon-Number selective phase gates},
  author = {Heeres, Reinier W. and Vlastakis, Brian and Holland, Eric and Krastanov, Stefan and Albert, Victor V. and Frunzio, Luigi and Jiang, Liang and Schoelkopf, Robert J.},
  journal = {Phys. Rev. Lett.},
  volume = {115},
  issue = {13},
  pages = {137002},
  numpages = {5},
  year = {2015},
  month = {Sep},
  publisher = {American Physical Society},
  doi = {10.1103/PhysRevLett.115.137002},
  url = {https://link.aps.org/doi/10.1103/PhysRevLett.115.137002}
}

@article{NatPhysDeng2024,
  author = {Deng, Xiaowei and Li, Sai and Chen, Zi-Jie and Ni, Zhongchu
            and Cai, Yanyan and Mai, Jiasheng and Zhang, Libo and Zheng, Pan
            and Yu, Haifeng and Zou, Chang-Ling and Liu, Song and Yan, Fei
            and Xu, Yuan and Yu, Dapeng},
  title = {Quantum-enhanced metrology with large {Fock} states},
  journal = {Nat. Phys.},
  volume = {20},
  pages = {1874--1880},
  year = {2024},
  doi = {10.1038/s41567-024-02619-5}
}

@article{RMPFrowis2018,
  title = {Macroscopic quantum states: Measures, fragility, and implementations},
  author = {Fr\"owis, Florian and Sekatski, Pavel and D\"ur, Wolfgang and Gisin, Nicolas and Sangouard, Nicolas},
  journal = {Rev. Mod. Phys.},
  volume = {90},
  issue = {2},
  pages = {025004},
  numpages = {52},
  year = {2018},
  month = {May},
  publisher = {American Physical Society},
  doi = {10.1103/RevModPhys.90.025004},
  url = {https://link.aps.org/doi/10.1103/RevModPhys.90.025004}
}

@article{npjQISturges2021,
  author = {Sturges, T. J. and McDermott, T. and Buraczewski, A. and Clements, W. R. and Renema, J. J. and Nam, S. W. and Gerrits, T. and Lita, A. and Kolthammer, W. S. and Eckstein, A. and Walmsley, I. A. and Stobi{\'n}ska, M.},
  title = {Quantum simulations with multiphoton {Fock} states},
  journal = {npj Quantum Inf.},
  volume = {7},
  pages = {91},
  year = {2021},
  doi = {10.1038/s41534-021-00427-w}
}

@article{PRLVogel1993,
  title = {Quantum state engineering of the radiation field},
  author = {Vogel, K. and Akulin, V. M. and Schleich, W. P.},
  journal = {Phys. Rev. Lett.},
  volume = {71},
  issue = {12},
  pages = {1816--1819},
  numpages = {0},
  year = {1993},
  month = {Sep},
  publisher = {American Physical Society},
  doi = {10.1103/PhysRevLett.71.1816},
  url = {https://link.aps.org/doi/10.1103/PhysRevLett.71.1816}
}

@article{PRLLaw1996,
  title = {Arbitrary control of a quantum electromagnetic field},
  author = {Law, C. K. and Eberly, J. H.},
  journal = {Phys. Rev. Lett.},
  volume = {76},
  issue = {7},
  pages = {1055--1058},
  numpages = {0},
  year = {1996},
  month = {Feb},
  publisher = {American Physical Society},
  doi = {10.1103/PhysRevLett.76.1055},
  url = {https://link.aps.org/doi/10.1103/PhysRevLett.76.1055}
}

@article{PRLBrattke2001,
  title = {Generation of photon number states on demand via cavity quantum electrodynamics},
  author = {Brattke, Simon and Varcoe, Benjamin T. H. and Walther, Herbert},
  journal = {Phys. Rev. Lett.},
  volume = {86},
  issue = {16},
  pages = {3534--3537},
  numpages = {0},
  year = {2001},
  month = {Apr},
  publisher = {American Physical Society},
  doi = {10.1103/PhysRevLett.86.3534},
  url = {https://link.aps.org/doi/10.1103/PhysRevLett.86.3534}
}

@article{PRLBrune1990,
  title = {Quantum nondemolition measurement of small photon numbers by {R}ydberg-atom phase-sensitive detection},
  author = {Brune, M. and Haroche, S. and Lefevre, V. and Raimond, J. M. and Zagury, N.},
  journal = {Phys. Rev. Lett.},
  volume = {65},
  issue = {8},
  pages = {976--979},
  numpages = {0},
  year = {1990},
  month = {Aug},
  publisher = {American Physical Society},
  doi = {10.1103/PhysRevLett.65.976},
  url = {https://link.aps.org/doi/10.1103/PhysRevLett.65.976}
}

@article{PRLFranifmmode2001,
  title = {Conditional large {F}ock state preparation and field state reconstruction in cavity {QED}},
  author = {Fran\ifmmode \mbox{\c{c}}\else \c{c}\fi{}a Santos, M. and Solano, E. and de Matos Filho, R. L.},
  journal = {Phys. Rev. Lett.},
  volume = {87},
  issue = {9},
  pages = {093601},
  numpages = {4},
  year = {2001},
  month = {Aug},
  publisher = {American Physical Society},
  doi = {10.1103/PhysRevLett.87.093601},
  url = {https://link.aps.org/doi/10.1103/PhysRevLett.87.093601}
}

@article{PRLUria2020,
  title = {Deterministic generation of large {F}ock states},
  author = {Uria, M. and Solano, P. and Hermann-Avigliano, C.},
  journal = {Phys. Rev. Lett.},
  volume = {125},
  issue = {9},
  pages = {093603},
  numpages = {6},
  year = {2020},
  month = {Aug},
  publisher = {American Physical Society},
  doi = {10.1103/PhysRevLett.125.093603},
  url = {https://link.aps.org/doi/10.1103/PhysRevLett.125.093603}
}

@article{NatureSayrin2011,
  author = {Sayrin, Cl{\'e}ment and Dotsenko, Igor and Zhou, Xingxing and Peaudecerf, Bruno and Rybarczyk, Th{\'e}o and Gleyzes, S{\'e}bastien and Rouchon, Pierre and Mirrahimi, Mazyar and Amini, Hadis and Brune, Michel and Raimond, Jean-Michel and Haroche, Serge},
  title = {Real-time quantum feedback prepares and stabilizes photon number states},
  journal = {Nature},
  volume = {477},
  pages = {73--77},
  year = {2011},
  doi = {10.1038/nature10376}
}

@article{NatureHofheinz2008,
  author = {Hofheinz, Max and Weig, E. M. and Ansmann, M. and Bialczak, Radoslaw C. and Lucero, Erik and Neeley, M. and O'Connell, A. D. and Wang, H. and Martinis, John M. and Cleland, A. N.},
  title = {Generation of {Fock} states in a superconducting quantum circuit},
  journal = {Nature},
  volume = {454},
  pages = {310--314},
  year = {2008},
  doi = {10.1038/nature07136}
}

@article{NatCommunPremaratne2017,
  author = {Premaratne, Shavindra P. and Wellstood, F. C. and Palmer, B. S.},
  title = {Microwave photon {Fock} state generation by stimulated {Raman} adiabatic passage},
  journal = {Nat. Commun.},
  volume = {8},
  pages = {14148},
  year = {2017},
  doi = {10.1038/ncomms14148}
}

@article{PRLWang2017,
  title = {Converting Quasiclassical States into Arbitrary Fock State Superpositions in a Superconducting Circuit},
  author = {Wang, W. and Hu, L. and Xu, Y. and Liu, K. and Ma, Y. and Zheng, Shi-Biao and Vijay, R. and Song, Y. P. and Duan, L.-M. and Sun, L.},
  journal = {Phys. Rev. Lett.},
  volume = {118},
  issue = {22},
  pages = {223604},
  numpages = {6},
  year = {2017},
  month = {Jun},
  publisher = {American Physical Society},
  doi = {10.1103/PhysRevLett.118.223604},
  url = {https://link.aps.org/doi/10.1103/PhysRevLett.118.223604}
}

@article{PRATan2014,
  title = {Deterministic quantum superpositions and Fock states of mechanical oscillators via quantum interference in single-photon cavity optomechanics},
  author = {Tan, Huatang},
  journal = {Phys. Rev. A},
  volume = {89},
  issue = {5},
  pages = {053829},
  numpages = {7},
  year = {2014},
  month = {May},
  publisher = {American Physical Society},
  doi = {10.1103/PhysRevA.89.053829},
  url = {https://link.aps.org/doi/10.1103/PhysRevA.89.053829}
}

@article{NatureChu2018,
  author = {Chu, Yiwen and Kharel, Prashanta and Yoon, Taekwan and Frunzio, Luigi and Rakich, Peter T. and Schoelkopf, Robert J.},
  title = {Creation and control of multi-phonon {Fock} states in a bulk acoustic-wave resonator},
  journal = {Nature},
  volume = {563},
  pages = {666--670},
  year = {2018},
  doi = {10.1038/s41586-018-0717-7}
}

@article{JMP1977Misra,
  author = {Misra, B. and Sudarshan, E. C. G.},
  title = {The {Zeno’s} paradox in quantum theory},
  journal = {J. Math. Phys.},
  volume = {18},
  number = {4},
  pages = {756-763},
  year = {1977},
  month = {04},
  issn = {0022-2488}, 
  doi = {10.1063/1.523304},
  url = {https://doi.org/10.1063/1.523304},
}

@article{PRL2003Nakazato,
  title = {Purification through {Z}eno-like measurements},
  author = {Nakazato, Hiromichi and Takazawa, Tomoko and Yuasa, Kazuya},
  journal = {Phys. Rev. Lett.},
  volume = {90},
  issue = {6},
  pages = {060401},
  numpages = {4},
  year = {2003},
  month = {Feb},
  publisher = {American Physical Society},
  doi = {10.1103/PhysRevLett.90.060401},
  url = {https://link.aps.org/doi/10.1103/PhysRevLett.90.060401}
}

@article{PRANakazato2004,
  title = {Preparation and entanglement purification of qubits through {Zeno}-like measurements},
  author = {Nakazato, Hiromichi and Unoki, Makoto and Yuasa, Kazuya},
  journal = {Phys. Rev. A},
  volume = {70},
  issue = {1},
  pages = {012303},
  numpages = {12},
  year = {2004},
  month = {Jul},
  publisher = {American Physical Society},
  doi = {10.1103/PhysRevA.70.012303},
  url = {https://link.aps.org/doi/10.1103/PhysRevA.70.012303}
}

@article{PRAZhang2026,
  title = {Efficient nonclassical state preparation via generalized parity measurement},
  author = {Zhang, Chen-yi and Jing, Jun},
  journal = {Phys. Rev. A},
  volume = {113},
  issue = {2},
  pages = {022420},
  numpages = {11},
  year = {2026},
  month = {Feb},
  publisher = {American Physical Society},
  doi = {10.1103/y4gs-7wt5},
  url = {https://link.aps.org/doi/10.1103/y4gs-7wt5}
}

@article{PRLZhou2012,
  title = {Field locked to a {Fock} state by quantum feedback with single photon corrections},
  author = {Zhou, X. and Dotsenko, I. and Peaudecerf, B. and Rybarczyk, T. and Sayrin, C. and Gleyzes, S. and Raimond, J. M. and Brune, M. and Haroche, S.},
  journal = {Phys. Rev. Lett.},
  volume = {108},
  issue = {24},
  pages = {243602},
  numpages = {5},
  year = {2012},
  month = {Jun},
  publisher = {American Physical Society},
  doi = {10.1103/PhysRevLett.108.243602},
  url = {https://link.aps.org/doi/10.1103/PhysRevLett.108.243602}
}

@article{arxivJin2026,
  title = {Deterministic generation of arbitrary {Fock} states via resonant subspace engineering},
  author = {Jin, Shan and Li, Ming and Cai, Weizhou and Chen, Zi-Jie and Xu, Yifang and Zhou, Yilong and Huang, Hongwei and Zhu, Yunlai and Hua, Ziyue and Guo, Guang-Can and Sun, Luyan and Wang, Xiaoting and Zou, Chang-Ling},
  eprint = {2602.12156},
  archivePrefix = {arXiv},
}

@article{arxivLi2026,
  title = {Scalable generation of macroscopic {Fock} states exceeding 10,000 photons},
  author = {Li, Ming and Cai, Weizhou and Hua, Ziyue and Xu, Yifang and Zhou, Yilong and Chen, Zi-Jie and Zou, Xu-Bo and Guo, Guang-Can and Sun, Luyan and Zou, Chang-Ling},
  eprint = {2601.05118},
  archivePrefix = {arXiv},
}

@article{PRAYan2023,
  title = {Generic eigenstate preparation via measurement-based purification},
  author = {Yan, Jia-shun and Jing, Jun},
  journal = {Phys. Rev. A},
  volume = {108},
  issue = {4},
  pages = {042215},
  numpages = {11},
  year = {2023},
  month = {Oct},
  publisher = {American Physical Society},
  doi = {10.1103/PhysRevA.108.042215},
  url = {https://link.aps.org/doi/10.1103/PhysRevA.108.042215}
}

@article{JPAPellonpaa2023,
  title = {{Naimark} dilations of qubit {POVMs} and joint measurements},
  author = {Pellonp{\"a}{\"a}, Juha-Pekka and Designolle, S{\'e}bastien and Uola, Roope},
  journal = {J. Phys. A: Math. Theor.},
  volume = {56},
  pages = {155303},
  year = {2023},
  doi = {10.1088/1751-8121/acc21c}
}

@article{PRLWiseman1995,
  title = {Adaptive phase measurements of optical modes: going beyond the marginal $Q$ distribution},
  author = {Wiseman, H. M.},
  journal = {Phys. Rev. Lett.},
  volume = {75},
  issue = {25},
  pages = {4587--4590},
  numpages = {0},
  year = {1995},
  month = {Dec},
  publisher = {American Physical Society},
  doi = {10.1103/PhysRevLett.75.4587},
  url = {https://link.aps.org/doi/10.1103/PhysRevLett.75.4587}
}

@article{PRLBerryWiseman2000,
  title = {Optimal states and almost optimal adaptive measurements for quantum interferometry},
  author = {Berry, D. W. and Wiseman, H. M.},
  journal = {Phys. Rev. Lett.},
  volume = {85},
  pages = {5098--5101},
  year = {2000},
  doi = {10.1103/PhysRevLett.85.5098}
}

@article{PRLArmen2002,
  title = {Adaptive homodyne measurement of optical phase},
  author = {Armen, Michael A. and Au, John K. and Stockton, John K. and Doherty, Andrew C. and Mabuchi, Hideo},
  journal = {Phys. Rev. Lett.},
  volume = {89},
  issue = {13},
  pages = {133602},
  numpages = {4},
  year = {2002},
  month = {Sep},
  publisher = {American Physical Society},
  doi = {10.1103/PhysRevLett.89.133602},
  url = {https://link.aps.org/doi/10.1103/PhysRevLett.89.133602}
}

@article{PRATsang2008,
  title = {Quantum theory of optical temporal phase and instantaneous frequency},
  author = {Tsang, Mankei and Shapiro, Jeffrey H. and Lloyd, Seth},
  journal = {Phys. Rev. A},
  volume = {78},
  issue = {5},
  pages = {053820},
  numpages = {15},
  year = {2008},
  month = {Nov},
  publisher = {American Physical Society},
  doi = {10.1103/PhysRevA.78.053820},
  url = {https://link.aps.org/doi/10.1103/PhysRevA.78.053820}
}

@article{PRLWheatley2010,
  title = {Adaptive optical phase estimation using time-symmetric quantum smoothing},
  author = {Wheatley, T. A. and Berry, D. W. and Yonezawa, H. and Nakane, D. and Arao, H. and Pope, D. T. and Ralph, T. C. and Wiseman, H. M. and Furusawa, A. and Huntington, E. H.},
  journal = {Phys. Rev. Lett.},
  volume = {104},
  issue = {9},
  pages = {093601},
  numpages = {4},
  year = {2010},
  month = {Mar},
  publisher = {American Physical Society},
  doi = {10.1103/PhysRevLett.104.093601},
  url = {https://link.aps.org/doi/10.1103/PhysRevLett.104.093601}
}

@article{ScienceYonezawa2012,
  title = {Quantum-enhanced optical-phase tracking},
  author = {Yonezawa, Hidehiro and Nakane, Daisuke and Wheatley, Trevor A. and Iwasawa, Kohjiro and Takeda, Shuntaro and Arao, Hajime and Ohki, Kentaro and Tsumura, Koji and Berry, Dominic W. and Ralph, Timothy C. and Wiseman, Howard M. and Huntington, Elanor H. and Furusawa, Akira},
  journal = {Science},
  volume = {337},
  pages = {1514--1517},
  year = {2012},
  doi = {10.1126/science.1225258}
}

@article{NatureRiste2013,
  title = {Deterministic entanglement of superconducting qubits by rarity measurement and feedback},
  author = {Rist{\`e}, D. and Dukalski, M. and Watson, C. A. and de Lange, G. and Tiggelman, M. J. and Blanter, Ya. M. and Lehnert, K. W. and Schouten, R. N. and DiCarlo, L.},
  journal = {Nature},
  volume = {502},
  pages = {350--354},
  year = {2013},
  doi = {10.1038/nature12513}
}

@article{NJPYuasa2009,
  title = {Efficient generation of a maximally entangled State by repeated on- and off-resonant scattering of ancilla qubits},
  author = {Yuasa, Kazuya and Burgarth, Daniel and Giovannetti, Vittorio and Nakazato, Hiromichi},
  journal = {New J. Phys.},
  volume = {11},
  pages = {123027},
  year = {2009},
  doi = {10.1088/1367-2630/11/12/123027}
}

@article{PRXQuantumSmith2024,
  title = {Constant-depth preparation of matrix product states with adaptive quantum circuits},
  author = {Smith, Kevin C. and Khan, Abid and Clark, Bryan K. and Girvin, S.M. and Wei, Tzu-Chieh},
  journal = {PRX Quantum},
  volume = {5},
  issue = {3},
  pages = {030344},
  numpages = {36},
  year = {2024},
  month = {Sep},
  publisher = {American Physical Society},
  doi = {10.1103/PRXQuantum.5.030344},
  url = {https://link.aps.org/doi/10.1103/PRXQuantum.5.030344}
}

@article{PRLSalvia2023,
  title = {Critical quantum metrology assisted by real-Time feedback control},
  author = {Salvia, Raffaele and Mehboudi, Mohammad and Perarnau-Llobet, Mart\'{\i}},
  journal = {Phys. Rev. Lett.},
  volume = {130},
  issue = {24},
  pages = {240803},
  numpages = {7},
  year = {2023},
  month = {Jun},
  publisher = {American Physical Society},
  doi = {10.1103/PhysRevLett.130.240803},
  url = {https://link.aps.org/doi/10.1103/PhysRevLett.130.240803}
}

@article{PRAWang2021,
  title = {Preparing Dicke states in a spin ensemble using phase estimation},
  author = {Wang, Yang and Terhal, Barbara M.},
  journal = {Phys. Rev. A},
  volume = {104},
  issue = {3},
  pages = {032407},
  numpages = {13},
  year = {2021},
  month = {Sep},
  publisher = {American Physical Society},
  doi = {10.1103/PhysRevA.104.032407},
  url = {https://link.aps.org/doi/10.1103/PhysRevA.104.032407}
}

@article{PRAZhang2024,
  title = {Generating Fock-state superpositions from coherent states by selective measurement},
  author = {Zhang, Chen-yi and Jing, Jun},
  journal = {Phys. Rev. A},
  volume = {110},
  issue = {4},
  pages = {042421},
  numpages = {12},
  year = {2024},
  month = {Oct},
  publisher = {American Physical Society},
  doi = {10.1103/PhysRevA.110.042421},
  url = {https://link.aps.org/doi/10.1103/PhysRevA.110.042421}
}

@article{PRATanaka2012,
  title = {Robust adaptive measurement scheme for qubit-state preparation},
  author = {Tanaka, Saki and Yamamoto, Naoki},
  journal = {Phys. Rev. A},
  volume = {86},
  issue = {6},
  pages = {062331},
  numpages = {6},
  year = {2012},
  month = {Dec},
  publisher = {American Physical Society},
  doi = {10.1103/PhysRevA.86.062331},
  url = {https://link.aps.org/doi/10.1103/PhysRevA.86.062331}
}

@article{PRLYu2026,
  title = {Efficient preparation of {Dicke} states},
  author = {Yu, Jeffery and Muleady, Sean R. and Wang, Yu-Xin and Schine, Nathan and Gorshkov, Alexey V. and Childs, Andrew M.},
  journal = {Phys. Rev. Lett.},
  volume = {136},
  issue = {3},
  pages = {030601},
  numpages = {9},
  year = {2026},
  month = {Jan},
  publisher = {American Physical Society},
  doi = {10.1103/9gjk-rgql},
  url = {https://link.aps.org/doi/10.1103/9gjk-rgql}
}

@article{PRBLi2011,
  title = {Nondeterministic ultrafast ground-state cooling of a mechanical resonator},
  author = {Li, Yong and Wu, Lian-Ao and Wang, Ying-Dan and Yang, Li-Ping},
  journal = {Phys. Rev. B},
  volume = {84},
  issue = {9},
  pages = {094502},
  numpages = {5},
  year = {2011},
  month = {Sep},
  publisher = {American Physical Society},
  doi = {10.1103/PhysRevB.84.094502},
  url = {https://link.aps.org/doi/10.1103/PhysRevB.84.094502}
}

@article{PRAYan2022,
  title = {Simultaneous cooling by measuring one ancillary system},
  author = {Yan, Jia-shun and Jing, Jun},
  journal = {Phys. Rev. A},
  volume = {105},
  issue = {5},
  pages = {052607},
  numpages = {10},
  year = {2022},
  month = {May},
  publisher = {American Physical Society},
  doi = {10.1103/PhysRevA.105.052607},
  url = {https://link.aps.org/doi/10.1103/PhysRevA.105.052607}
}

@article{PRAJin2022,
  title = {Measurement-induced nuclear spin polarization},
  author = {Jin, Zhu-yao and Yan, Jia-shun and Jing, Jun},
  journal = {Phys. Rev. A},
  volume = {106},
  issue = {6},
  pages = {062605},
  numpages = {10},
  year = {2022},
  month = {Dec},
  publisher = {American Physical Society},
  doi = {10.1103/PhysRevA.106.062605},
  url = {https://link.aps.org/doi/10.1103/PhysRevA.106.062605}
}

@article{PRAKonar2022,
  title = {Refrigeration via purification through repeated measurements},
  author = {Konar, Tanoy Kanti and Ghosh, Srijon and Sen(De), Aditi},
  journal = {Phys. Rev. A},
  volume = {106},
  issue = {2},
  pages = {022616},
  numpages = {11},
  year = {2022},
  month = {Aug},
  publisher = {American Physical Society},
  doi = {10.1103/PhysRevA.106.022616},
  url = {https://link.aps.org/doi/10.1103/PhysRevA.106.022616}
}

@article{PRAppliedYan2023,
  title = {Charging by Quantum Measurement},
  author = {Yan, Jia-shun and Jing, Jun},
  journal = {Phys. Rev. Appl.},
  volume = {19},
  issue = {6},
  pages = {064069},
  numpages = {11},
  year = {2023},
  month = {Jun},
  publisher = {American Physical Society},
  doi = {10.1103/PhysRevApplied.19.064069},
  url = {https://link.aps.org/doi/10.1103/PhysRevApplied.19.064069}
}

@article{PRATinggui2024,
  title = {Local-projective-measurement-enhanced quantum battery capacity},
  author = {Zhang, Tinggui and Yang, Hong and Fei, Shao-Ming},
  journal = {Phys. Rev. A},
  volume = {109},
  issue = {4},
  pages = {042424},
  numpages = {6},
  year = {2024},
  month = {Apr},
  publisher = {American Physical Society},
  doi = {10.1103/PhysRevA.109.042424},
  url = {https://link.aps.org/doi/10.1103/PhysRevA.109.042424}
}

@article{PRAWu2004,
  title = {Long-range entanglement generation via frequent measurements},
  author = {Wu, L.-A. and Lidar, D. A. and Schneider, S.},
  journal = {Phys. Rev. A},
  volume = {70},
  issue = {3},
  pages = {032322},
  numpages = {7},
  year = {2004},
  month = {Sep},
  publisher = {American Physical Society},
  doi = {10.1103/PhysRevA.70.032322},
  url = {https://link.aps.org/doi/10.1103/PhysRevA.70.032322}
}

@article{PRAUys2007,
  title = {Quantum states for {Heisenberg}-limited interferometry},
  author = {Uys, H. and Meystre, P.},
  journal = {Phys. Rev. A},
  volume = {76},
  issue = {1},
  pages = {013804},
  numpages = {9},
  year = {2007},
  month = {Jul},
  publisher = {American Physical Society},
  doi = {10.1103/PhysRevA.76.013804},
  url = {https://link.aps.org/doi/10.1103/PhysRevA.76.013804}
}

@article{PRLHolland1993,
  title = {Interferometric detection of optical phase shifts at the {Heisenberg} limit},
  author = {Holland, M. J. and Burnett, K.},
  journal = {Phys. Rev. Lett.},
  volume = {71},
  issue = {9},
  pages = {1355--1358},
  numpages = {0},
  year = {1993},
  month = {Aug},
  publisher = {American Physical Society},
  doi = {10.1103/PhysRevLett.71.1355},
  url = {https://link.aps.org/doi/10.1103/PhysRevLett.71.1355}
}

@article{PRLDescamps2024,
  title = {Superselection Rules and Bosonic Quantum Computational Resources},
  author = {Descamps, Eloi and Fabre, Nicolas and Saharyan, Astghik and Keller, Arne and Milman, P{\'e}rola},
  journal = {Phys. Rev. Lett.},
  volume = {133},
  pages = {260605},
  year = {2024},
  doi = {10.1103/PhysRevLett.133.260605}
}

@article{OpticaQDescamps2026,
  title = {Unified Framework for Bosonic Quantum Information Encoding, Resources, and Universality from Superselection Rules},
  author = {Descamps, Eloi and Saharyan, Astghik and Chivet, Adrien and Keller, Arne and Milman, P{\'e}rola},
  journal = {Optica Quantum},
  volume = {4},
  number = {2},
  pages = {148--161},
  year = {2026},
  doi = {10.1364/OPTICAQ.581218}
}

@article{Xiong2026LargeFock,
  title = {Deterministic and Scalable Generation of Large Fock States},
  author = {Xiong, Mo and Han, Jize and Cao, Chuanzhen and Li, Jinbin and Liu, Qi and Huang, Zhiguo and Xue, Ming},
  eprint = {2601.10559},
  archivePrefix = {arXiv},
}

@article{PRLJeffrey2014,
  title = {Fast accurate state measurement with superconducting qubits},
  author = {Jeffrey, Evan and Sank, Daniel and Mutus, J. Y. and White, T. C. and Kelly, J. and Barends, R. and Chen, Y. and Chen, Z. and Chiaro, B. and Dunsworth, A. and Megrant, A. and O'Malley, P. J. J. and Neill, C. and Roushan, P. and Vainsencher, A. and Wenner, J. and Cleland, A. N. and Martinis, John M.},
  journal = {Phys. Rev. Lett.},
  volume = {112},
  issue = {19},
  pages = {190504},
  numpages = {5},
  year = {2014},
  month = {May},
  publisher = {American Physical Society},
  doi = {10.1103/PhysRevLett.112.190504},
  url = {https://link.aps.org/doi/10.1103/PhysRevLett.112.190504}
}

@article{PRAppliedWalter2017,
  title = {Rapid high-fidelity single-shot dispersive readout of superconducting qubits},
  author = {Walter, T. and Kurpiers, P. and Gasparinetti, S. and Magnard, P. and Poto\ifmmode \check{c}\else \v{c}\fi{}nik, A. and Salath\'e, Y. and Pechal, M. and Mondal, M. and Oppliger, M. and Eichler, C. and Wallraff, A.},
  journal = {Phys. Rev. Appl.},
  volume = {7},
  issue = {5},
  pages = {054020},
  numpages = {11},
  year = {2017},
  month = {May},
  publisher = {American Physical Society},
  doi = {10.1103/PhysRevApplied.7.054020},
  url = {https://link.aps.org/doi/10.1103/PhysRevApplied.7.054020}
}

@article{PRAppliedHeinsoo2018,
  title = {Rapid High-fidelity Multiplexed Readout of Superconducting Qubits},
  author = {Heinsoo, Johannes and Andersen, Christian Kraglund and Remm, Ants and Krinner, Sebastian and Walter, Theodore and Salath\'e, Yves and Gasparinetti, Simone and Besse, Jean-Claude and Poto\ifmmode \check{c}\else \v{c}\fi{}nik, Anton and Wallraff, Andreas and Eichler, Christopher},
  journal = {Phys. Rev. Appl.},
  volume = {10},
  issue = {3},
  pages = {034040},
  numpages = {14},
  year = {2018},
  month = {Sep},
  publisher = {American Physical Society},
  doi = {10.1103/PhysRevApplied.10.034040},
  url = {https://link.aps.org/doi/10.1103/PhysRevApplied.10.034040}
}

@article{PRXQMarxer2026,
  title = {Above 99.9\% fidelity single-qubit gates, two-qubit gates, and readout in a single superconducting quantum device},
  author = {Marxer, Fabian and Mro\ifmmode \dot{z}\else \.{z}\fi{}ek, Jakub and Andersson, Joona and Abdurakhimov, Leonid and Adam, Janos and Bergholm, Ville and Beriwal, Rohit and Chan, Chun Fai and Dahl, Saga and Das, Soumya Ranjan and Deppe, Frank and Fedorets, Olexiy and Gao, Zheming and Gomez Frieiro, Alejandro and Gusenkova, Daria and Guthrie, Andrew and Hiltunen, Tuukka and Hsu, Hao and Hyypp\"a, Eric and Ikonen, Joni and Inel, Sinan and Jolin, Shan W. and Karis, Azad and Kim, Seung-Goo and Kindel, William and Komlev, Anton and Koistinen, Miikka and Kokkoniemi, Roope and Kumar, Snigdha and Ku, Hsiang-Sheng and Lamprich, Julia and Laine, Sami and Landra, Alessandro and Lee, Lan-Hsuan and Lethif, Nizar and Liebermann, Per and Liu, Wei and Mitra, Kunal and Myll\"ari, Tuomas and Ockeloen-Korppi, Caspar and Orell, Tuure and Plyshch, Alexander and R\"abin\"a, Jukka and Rebello, Arthur and Renger, Michael and Reentil\"a, Outi and Ritvas, Jussi and Saarinen, Sampo and Salmenkivi, Otto and Sarsby, Matthew and Savytskyi, Mykhailo and Selinmaa, Ville and Steggles, Matthew and Takala, Eelis and Takmakov, Ivan and Tarasinski, Brian and Tuorila, Jani and V\"alimaa, Alpo and Verjauw, Jeroen and Wesdorp, Jaap and Wurz, Nicola and Qiu, Wei and Zhu, Lihuang and Hassel, Juha and Heinsoo, Johannes and Geresdi, Attila and Veps\"al\"ainen, Antti},
  journal = {PRX Quantum},
  volume = {7},
  issue = {2},
  pages = {020333},
  numpages = {38},
  year = {2026},
  month = {May},
  publisher = {American Physical Society},
  doi = {10.1103/n86s-2b88},
  url = {https://link.aps.org/doi/10.1103/n86s-2b88}
}

@book{NielsenChuang2010, 
place={Cambridge}, 
title={Quantum Computation and Quantum Information: 10th Anniversary Edition}, 
publisher={Cambridge University Press}, 
author={Nielsen, Michael A. and Chuang, Isaac L.}, 
year={2010}}

@article{PRL2014Asadian,
  title = {Probing macroscopic realism via {Ramsey} correlation measurements},
  author = {Asadian, A. and Brukner, C. and Rabl, P.},
  journal = {Phys. Rev. Lett.},
  volume = {112},
  issue = {19},
  pages = {190402},
  numpages = {5},
  year = {2014},
  month = {May},
  publisher = {American Physical Society},
  doi = {10.1103/PhysRevLett.112.190402},
  url = {https://link.aps.org/doi/10.1103/PhysRevLett.112.190402}
}

@article{PRX2018Fluhmann,
  title = {Sequential modular position and momentum measurements of a trapped ion mechanical oscillator},
  author = {Fl\"uhmann, C. and Negnevitsky, V. and Marinelli, M. and Home, J. P.},
  journal = {Phys. Rev. X},
  volume = {8},
  issue = {2},
  pages = {021001},
  numpages = {17},
  year = {2018},
  month = {Apr},
  publisher = {American Physical Society},
  doi = {10.1103/PhysRevX.8.021001},
  url = {https://link.aps.org/doi/10.1103/PhysRevX.8.021001}
}

\end{document}